\documentclass[11pt,a4paper]{article}
\pdfoutput=1
\usepackage{jheppub}
\usepackage{tikz}
\usetikzlibrary{intersections, calc, arrows.meta}
\usepackage{subcaption}
\usepackage{bm}

\DeclareMathOperator{\Tr}{Tr}

\newcommand{\ri}{\mathrm{i}}

\newcommand{\cob}{\delta}

\newcommand{\hf}{\frac{1}{2}}

\newcommand{\til}[1]{\widetilde{#1}}

\newcommand{\Si}{\Sigma}
\renewcommand{\b}[1]{\overline{#1}}
\newcommand{\del}{\partial}

\newcommand{\bra}{\langle}
\newcommand{\ket}{\rangle}
\newcommand{\la}{\lambda}

\newcommand{\h}[1]{\widehat{#1}}
\newcommand{\bt}{\beta}

\newcommand{\rt}[1]{\sqrt{#1}}

\newcommand{\cI}{\mathcal{I}}
\newcommand{\cE}{\mathcal{E}}

\newcommand{\bbZ}{{\mathbb Z}}
\newcommand{\gs}{g_{\rm s}}
\newcommand{\tS}{\widetilde{S}}
\newcommand{\tE}{\widetilde{E}}
\newcommand{\tv}{\tilde{v}}
\newcommand{\tc}{t_{\rm c}}
\newcommand{\Ec}{E_0^{\rm c}}

\begin{document}

\title{Page curve from dynamical branes in JT gravity}

\author[a]{Kazumi Okuyama}
\author[b]{and Kazuhiro Sakai}

\affiliation[a]{Department of Physics, Shinshu University,\\
3-1-1 Asahi, Matsumoto 390-8621, Japan}
\affiliation[b]{Institute of Physics, Meiji Gakuin University,\\
1518 Kamikurata-cho, Totsuka-ku, Yokohama 244-8539, Japan}

\emailAdd{kazumi@azusa.shinshu-u.ac.jp, kzhrsakai@gmail.com}

\abstract{
We study the Page curve of an evaporating black hole
using a toy model given by Jackiw-Teitelboim gravity
with Fateev-Zamolodchikov-Zamolodchikov-Teschner (FZZT) antibranes.
We treat the anti-FZZT branes as dynamical objects, taking
their back-reaction into account.
We construct the entanglement entropy from the dual matrix model
and study its behavior as a function of
the 't Hooft coupling $t$ proportional to the number of branes,
which plays the role of time.
By numerical computation we observe that
the entropy first increases and then decreases as $t$ grows,
reproducing the well-known behavior of the Page curve of
an evaporating black hole.
The system finally exhibits a phase transition,
which may be viewed as the end of the evaporation.
We study the critical behavior of the entropy near the phase transition.
We also make a conjecture about
the late-time monotonically decreasing behavior of the entropy.
We prove it in a certain limit
as well as give an intuitive explanation
by means of the dual matrix model.}

\maketitle

\section{Introduction}

The black hole information paradox has been a long-standing puzzle
in the study of quantum gravity \cite{Hawking:1976ra}.
In particular, the growing behavior of the entropy of
thermal radiation based on Hawking's calculation
\cite{Hawking:1975vcx}
apparently contradicts with the unitarity of the quantum mechanics
which requires that the black hole stays in a pure state.
For an evaporating black hole,
the Page curve \cite{Page:1993wv},
a plot of the entanglement entropy of
the Hawking radiation as a function of time,
should show a decreasing behavior toward the end of evaporation.
Recent studies revealed that
the gravitational path integral receives, even semi-classically,
contributions from saddle-points other than
the classical black hole solution, namely the replica wormholes
\cite{Penington:2019npb,Almheiri:2019psf}.
This is a key to understand how the Page curve
is obtained in an expected form,
which partly resolves the information paradox.
The idea was refined in the form of the
island formula \cite{Almheiri:2019hni},
which was first derived by means of holography
\cite{Ryu:2006bv,Hubeny:2007xt,
Lewkowycz:2013nqa,Barrella:2013wja,
Faulkner:2013ana,Engelhardt:2014gca,
Penington:2019npb,Almheiri:2019psf,Almheiri:2019hni}
and then consolidated by directly evaluating the gravitational
path integral in quantum gravity in two dimensions
\cite{Penington:2019kki,Almheiri:2019qdq}.
See \cite{Almheiri:2020cfm} for a recent review and references therein.

In \cite{Penington:2019kki} the Page curve was studied by using
Jackiw-Teitelboim (JT) gravity \cite{Jackiw:1984je,Teitelboim:1983ux}
with the end-of-the-world (EOW) branes.\footnote{A classification of
branes in JT gravity is found in \cite{Goel:2020yxl}.}
Roughly speaking, the system is viewed as a generalization of
the original Page's model \cite{Page:1993df} (see
\cite{Okuyama:2021ylc} for recent exact results).
Page's calculation starts with
a random pure state in the bipartite Hilbert space
consisting of two subspaces that
represent the interior and exterior of a black hole.
Taking ensemble average of the state in either of the subspaces,
one obtains the reduced density matrix, from which
the entanglement entropy is calculated.
In Page's model the ensemble is Gaussian in both subspaces.
In the case of JT gravity with EOW branes,
the ensemble in the interior is Gaussian
whereas the average in the exterior is described by
the double-scaled matrix integral of JT gravity \cite{Saad:2019lba}.
The size of the interior subspace, which is identified with
the number of branes, plays the role of time.

In this paper we propose another simple toy model
to understand the Page curve: JT gravity with
Fateev-Zamolodchikov-Zamolodchikov-Teschner (FZZT) antibranes
\cite{Fateev:2000ik,Teschner:2000md}.
Our model is a simplified variant of the model of
\cite{Penington:2019kki},
with the EOW branes replaced by anti-FZZT branes.
In our previous paper \cite{Okuyama:2021eju} we showed that
the matrix model description of the EOW brane in \cite{Gao:2021uro}
corresponds to that of a collection of infinitely many
anti-FZZT branes with a particular set of parameters.
It is therefore simpler to consider JT gravity with a single
kind of anti-FZZT branes.

Despite this simplification, our model captures several features
of black hole entropy. Most notably, 
the entanglement entropy exhibits the late-time decreasing behavior
which is characteristic of an evaporating black hole.\footnote{The
Page curve of an evaporating black hole in JT gravity was studied
in different approaches \cite{Goto:2020wnk,Cadoni:2021ypx}.}
To reproduce this decreasing behavior,
it is crucial to treat branes as dynamical objects.
In the previous studies,
branes are treated as either dynamical \cite{Gao:2021uro}
or non-dynamical \cite{Penington:2019kki}.
We will see from numerical computation that
the late-time decreasing behavior of the entropy is reproduced
only when we treat anti-FZZT branes as dynamical objects.
In fact, we consider the 't~Hooft limit, in which
the back-reaction of branes is not negligible
and one has to treat branes as dynamical objects.
We will also study how this decreasing behavior
arises from the viewpoint of the matrix model
and make a conjecture about monotonicity,
which we will prove in a certain limit.

Our model exhibits a phase transition as ``time'' grows.
The transition may be viewed as the end of the evaporation of
black hole.
We will study the critical behavior of the entropy
near the transition point.
One can consider JT gravity with FZZT branes
and study the Page curve in the same manner.
In this case, however, no phase transition occurs
and the entropy continue increasing.
All these results derived from the matrix model
are in perfect accordance with
the semi-classical analysis on the gravity side:
Dilaton gravities with nontrivial dilaton potential
were studied as deformations of JT gravity
\cite{Maxfield:2020ale,Witten:2020wvy}
and black hole solutions
in these gravities were also studied \cite{Witten:2020ert}.
JT gravity with (anti-)FZZT branes can be viewed as
this type of dilaton gravity \cite{Okuyama:2021eju}.
We will study its black hole solutions
and see the continuous growth of the entropy
in the FZZT setup and the phase transition in the anti-FZZT setup.
Thus, in this paper we concentrate on the case of anti-FZZT branes.

This paper is organized as follows.
In section~\ref{sec:setup},
we describe our model and explain the general
method of computing the entropy of the Hawking radiation.
In section~\ref{sec:phase},
we explain how the phase transition occurs
and study the critical behavior of the entropy.
In section~\ref{sec:page},
we numerically study the Page curve,
i.e.~the evolution of the entropy as a function of
the 't Hooft coupling.
We also make a conjecture about the late-time monotonically
decreasing behavior of the entropy.
In section~\ref{sec:mono},
we prove the conjecture in a certain limit.
We also give an intuitive explanation of the reason
why the entropy decreases.
In section~\ref{sec:dilaton},
we study black hole solutions
from the viewpoint of dilaton gravity.
Finally we conclude in section~\ref{sec:conclusion}.
In appendix~\ref{app:SDeq},
we give a derivation of the Schwinger-Dyson equation
\eqref{eq:SD-R} based on the saddle point method.

\section{Entropy of radiation from dynamical anti-FZZT branes}
\label{sec:setup}

In this section we will describe our model and explain the general
method of computing the entropy of the Hawking radiation.
In many parts of our formulation
we follow the method of \cite{Penington:2019kki}
with EOW branes being replaced by anti-FZZT branes.
In our study, however, branes are treated as dynamical objects.
This is along the lines of \cite{Gao:2021uro}
and an important difference from \cite{Penington:2019kki}.

\subsection{Matrix integral and black hole microstates}

Let us consider general 2d topological gravity with $K$ dynamical
anti-FZZT branes.\footnote{In this paper we will eventually
restrict ourselves to the JT gravity case,
but most parts of our formalism can be applied to other 2d gravities
as well.}
It is described by the double scaling limit
of the matrix integral
\begin{equation}
\begin{aligned}
  Z&=\int dH e^{-\Tr V(H)}\det(\xi+H)^{-K}\\
&=
\int dH dQdQ^\dag e^{-\Tr V(H)-\Tr Q^\dag (\xi+H)Q }.
\end{aligned} 
\end{equation}
Here $H$ and $Q$ are $N\times N$ hermitian
and $N\times K$ complex matrices respectively.
$\xi$ is a parameter characterizing the anti-FZZT brane,
which is now taken to be common to all $K$ branes.
The potential could have been normalized as
\begin{align}
\frac{1}{\gs}V(H),
\end{align}
where $\gs$ is the genus counting parameter,
so that the genus expansion is manifest.
In this paper we include $\gs^{-1}$ in $V$ for simplicity.
In the double scaling limit, $N$ is sent to infinity
and the potential turns into the effective potential.
In this paper we will further take the 't Hooft limit
\begin{align}
K\to\infty,~\gs\to 0\quad\text{with}~~ t\equiv \gs K~~\text{fixed}
\label{eq:tHooft}
\end{align}
and evaluate quantities in the planar approximation.
That is, we will ignore all higher-order corrections
of expansions in $\gs$ and $K^{-1}$.

The matrices $H,Q$ are often denoted
by their components $H_{ab}$, $Q_{ai}$,
where $a,b=1,\ldots,N$ are ``color'' indices
and $i,j=1,\ldots,K$ are ``flavor'' indices.
The color degrees of freedom are used for describing bulk gravity
while the flavor degrees of freedom
are thought of as describing the interior partners of
the early Hawking radiation.
One can regard the matrix element $H_{ab}$ as
\begin{align}
H_{ab}=\bra a|H|b\ket,
\end{align}
where $H$ is a Hamiltonian operator and $\{|a\ket\}_{a=1}^N$
form an orthonormal basis
of the corresponding $N$ dimensional Hilbert space
\begin{align}
\bra a|b\ket =\delta_{ab},\qquad 1=\sum_a |a\ket\bra a|.
\end{align}
For $i$th random vector variable $Q_{ai}$
we consider the (canonical) thermal pure quantum state
\cite{Sugiura:2013pla,Goto:2021mbt} 
\begin{equation}
\begin{aligned}
 |\psi_i\ket=\sum_{a}e^{-\hf\bt H}|a\ket Q_{ai}
 =\sum_{a,b}|b\ket(e^{-\hf\bt H})_{ba}Q_{ai}.
\end{aligned} 
\label{eq:TPQ}
\end{equation}
Here $\beta$ is the inverse temperature, which is identified with
the (renormalized) length of an asymptotic boundary in 2d gravity.
$|\psi_i\ket$ play the role of the black hole microstates.

\subsection{Ensemble average}

To study the entropy, we will compute
the average of overlaps such as $\bra\psi_i|\psi_j\ket$.
We define the average of ${\cal O}$ by
\begin{equation}
\bra\overline{{\cal O}}\ket
=\int dH dQdQ^\dag e^{-\Tr V(H)-\Tr Q^\dag (\xi+H)Q }{\cal O}.
\end{equation}
Here the angle brackets $\bra\ \ket$ represent
averaging over the color degrees of freedom
while the overline $\overline{\rule{1em}{0ex}\rule{0em}{1.5ex}}$
represents averaging over the flavor degrees of freedom.
It is convenient to change the variable as
\begin{equation}
\begin{aligned}
 Q=(\xi+H)^{-\hf}C,
\end{aligned} 
\end{equation}
so that the new random variable $C$ obeys the Gaussian distribution
\begin{equation}
\bra\overline{{\cal O}}\ket
=\int dH dCdC^\dag\det(\xi+H)^{-K} e^{-\Tr V(H)-\Tr C^\dag C} {\cal O}.
\label{eq:vev}
\end{equation}
Thus,
in terms of $C$ the flavor average becomes nothing
but the Gaussian average.
Note that the determinant factor is recovered from
the integration measure.
On the other hand, the thermal pure quantum state \eqref{eq:TPQ} becomes
(see also appendix D of \cite{Penington:2019kki})
\begin{equation}
\begin{aligned}
 |\psi_i\ket=\sum_{a,b}|b\ket
 \bigl[e^{-\hf\bt H}(\xi+H)^{-\hf}\bigr]_{ba}C_{ai}.
\end{aligned} 
\label{eq:psi-antiFZZT}
\end{equation}
For our discussion it is convenient to
express \eqref{eq:psi-antiFZZT} as
\begin{equation}
\begin{aligned}
 |\psi_i\ket=\sum_{a,b}|b\ket (\rt{A})_{ba}C_{ai}
\end{aligned} 
\end{equation}
with
\begin{equation}
\begin{aligned}
A(H)=\frac{e^{-\bt H}}{\xi+H}.
\end{aligned}
\end{equation}
We then consider the overlaps such as
\begin{equation}
\begin{aligned}
W_{ij}&\equiv
 \bra\psi_i|\psi_j\ket=\sum_{a,b}A_{ab}C_{ai}^*C_{bj},\\
W_{ij}W_{ji}&=
|\bra\psi_i|\psi_j\ket|^2
 =\sum_{a,b,a',b'}A_{ab}A_{b'a'}C_{ai}^*C_{bj}C_{a'i}C_{b'j}^*.
\end{aligned} 
\label{eq:overlaps}
\end{equation}
Recall that
the Gaussian average of $C$ can be computed by the Wick contraction
\begin{equation}
\begin{aligned}
 \b{C_{ai}^*C_{bj}}&=\cob_{ab}\cob_{ij},\\
\b{C_{ai}^*C_{bj}C_{a'i}C_{b'j}^*}&=\cob_{ij}\cob_{ab}\cob_{a'b'}
+\cob_{aa'}\cob_{bb'}.
\end{aligned} 
\label{eq:C-ave}
\end{equation}
By using these formulas, the average of the overlaps \eqref{eq:overlaps}
are given by
\begin{equation}
\begin{aligned}
 \b{\bra\psi_i|\psi_j\ket}&=\cob_{ij}\Tr A,\\
\b{|\bra\psi_i|\psi_j\ket|^2}&=\cob_{ij}(\Tr A)^2+\Tr A^2.
\end{aligned} 
\label{eq:psi-average}
\end{equation}

As discussed in \cite{Penington:2019kki}, one can 
visualize the above computation \eqref{eq:psi-average}
by drawing diagrams. For instance, 
$\bra\psi_i|\psi_j\ket$ in \eqref{eq:overlaps}
can be represented by the following diagram
\begin{equation}
\begin{aligned}
\begin{tikzpicture}[scale=0.4]
\draw (0,0) circle [x radius=0.1, y radius=0.1];
\draw [dashed] (0.1,0) -- (4,0); 
\draw [thick,-]  (4,0) .. controls (6,1) and (9,-1) .. (10,0);
\draw [dashed] (10,0) -- (14.1,0);
\draw (14.1,0) circle [x radius=0.1, y radius=0.1];
\draw (-1,0) node [left]{$\bra\psi_i|\psi_j\ket=(C^\dagger AC)_{ij}=$};
\draw (0,0) node [above]{$i$};
\draw (14.1,0) node [above]{$j$};
\draw (2,0) node [above]{$(C^\dagger)_{ia}$};
\draw (7,0) node [above]{$A_{ab}$};
\draw (12,0) node [above]{$C_{bj}$};
\end{tikzpicture}.
\end{aligned} 
\end{equation}
The black thick curve labeled by the color matrix $A_{ab}$
corresponds to the asymptotic boundary of 2d spacetime while the
dashed lines correspond to the flavor degrees of freedom $C, C^\dagger$.
The gravitational path integral in the presence of branes
is given by the matrix integral \eqref{eq:vev}.
One can easily see that the gravitational computations in
eq.~(2.10) and 
Figure 3 of \cite{Penington:2019kki}
agree with the first and the second lines of \eqref{eq:psi-average}, 
respectively.

\subsection{Reduced density matrix of radiation}

As explained in \cite{Penington:2019kki},
the reduced density matrix of radiation is
represented by the ensemble average of
\begin{equation}
\begin{aligned}
 \varrho_{ij}=\frac{W_{ij}}{\sum_{i=1}^K W_{ii}}
 =\left(\frac{W}{\Tr W}\right)_{ij}.
\end{aligned} 
\end{equation}
This is normalized as $\Tr\varrho=1$.
Let us first consider the ``purity'' $\b{\Tr\varrho^2}$
as an example.
In the planar approximation,
we can take the average of the numerator and the denominator
of $\Tr\varrho^2$ independently
\begin{equation}
\begin{aligned}
 \b{\Tr\varrho^2}\approx 
\frac{\overline{\Tr W^2}}{\overline{\Tr W}^2}
&=\frac{K(\Tr A)^2+K^2\Tr A^2}{(K\Tr A)^2}\\
&=\frac{1}{K}+\frac{\Tr A^2}{(\Tr A)^2}.
\end{aligned}
\end{equation}
Similarly,
the average of $\Tr\varrho^n$ is approximated as
\begin{equation}
\begin{aligned}
 \b{\Tr\varrho^n}\approx\frac{\b{\Tr W^n}}{(\b{\Tr W})^n}
 =\frac{\b{\Tr W^n}}{(K\Tr A)^n}.
\end{aligned} 
\end{equation}
$\b{\Tr W^n}$ in the numerator can be computed by
using the Wick contraction of $C$ and $C^\dag$.
In the planar approximation one obtains
\begin{equation}
\begin{aligned}
 \b{\Tr W}&=K\Tr A,\\
\b{\Tr W^2}&=K(\Tr A)^2+K^2\Tr A^2,\\
\b{\Tr W^3}&=K(\Tr A)^3
 +3K^2\Tr A^2\Tr A+K^3\Tr A^3,\\
\b{\Tr W^4}&=K(\Tr A)^4+6K^2\Tr A^2(\Tr A)^2+2K^3(\Tr A^2)^2
 +4K^3\Tr A^3\Tr A+K^4\Tr A^4.
\end{aligned} 
\label{eq:bTrWn}
\end{equation}
In fact, $\b{\Tr W^n}$ can be computed efficiently by means of
the generating function
\begin{equation}
\begin{aligned}
 R(\la)=\Tr\frac{1}{\la-\varrho}
 =\sum_{n=0}^\infty \frac{\Tr\varrho^n}{\la^{n+1}}
 =\frac{K}{\la}
  +\sum_{n=1}^\infty\frac{\b{\Tr W^n}}{\la^{n+1}(K\Tr A)^n}.
\end{aligned} 
\label{eq:Rexp}
\end{equation}
In the planar approximation $R(\la)$ satisfies
\begin{equation}
\begin{aligned}
 \la R(\la)=K+\sum_{n=1}^\infty \frac{R(\la)^n\Tr A^n}{(K\Tr A)^n}.
\end{aligned}
\label{eq:SD-R} 
\end{equation}
This equation was derived diagrammatically in \cite{Penington:2019kki}.
We give an alternative derivation
based on the saddle point method
in appendix~\ref{app:SDeq}.
By plugging \eqref{eq:Rexp} into \eqref{eq:SD-R},
$\b{\Tr W^n}$ can be obtained recursively.
In this way, in the planar approximation we obtain
\begin{align}
\bra\b{\Tr\varrho^n}\ket\approx
 \frac{1}{K^{n-1}}+\frac{n(n-1)}{2K^{n-2}}
 \frac{\bra\Tr A^2\ket}{\bra\Tr A\ket^2}
 +\cdots
 +\frac{n}{K}\frac{\bra\Tr A^{n-1}\ket}{\bra\Tr A\ket^{n-1}}
 +\frac{\bra\Tr A^n\ket}{\bra\Tr A\ket^n}.
\label{eq:Tr-rho^n}
\end{align}

Thus, to compute $\bra\b{\Tr\varrho^n}\ket$
we need to evaluate
\begin{align}
\bra\Tr A^n\ket
 =\int dH e^{-\Tr V(H)} \det(\xi+H)^{-K} \Tr A^n.
\label{eq:TrAn}
\end{align}
We evaluate it in the double scaling limit.
In the planar approximation
we have only to consider the genus zero part.
It can be expressed in terms of
the leading-order density $\rho_0(E)$
of the eigenvalues of $H$.
As we studied in \cite{Okuyama:2021eju},
for Witten-Kontsevich topological gravity with general couplings
$\{t_k\}\ (k\in\bbZ_{\ge 0})$, $\rho_0(E)$ is given by
\begin{align}
\rho_0(E)
 =\frac{1}{\sqrt{2}\pi\gs}\int_{E_0}^E\frac{dv}{\sqrt{E-v}}
  \frac{\partial f(-v)}{\partial (-v)}
\label{eq:rho0}
\end{align}
with
\begin{align}
f(u):=\sum_{k=0}^\infty(\delta_{k,1}-t_k)\frac{u^k}{k!}.
\label{eq:f}
\end{align}
The threshold energy $E_0$ is determined by the condition
(the genus-zero string equation)
\begin{align}
f(-E_0)=0.
\label{eq:E0cond}
\end{align}

In this paper we consider JT gravity,
which corresponds to a particular background $t_k=\gamma_k$ with
\cite{Mulase:2006baa,Dijkgraaf:2018vnm,Okuyama:2019xbv}\footnote{Another
way to obtain JT gravity is to take the $p\to\infty$ limit of
the $(2,p)$ minimal string \cite{Seiberg:2019,Saad:2019lba}.
Entanglement entropy in this context was studied
recently in \cite{Hirano:2021rzg}.}
\begin{align}
\gamma_0=\gamma_1=0,\quad \gamma_k=\frac{(-1)^k}{(k-1)!}\quad (k\ge 2).
\end{align}
As we studied in \cite{Okuyama:2021eju},
the effect of anti-FZZT branes, i.e.~the insertion of $\det(\xi+H)^{-K}$
amounts to shifting the couplings $t_k$ of topological gravity
as\footnote{This shift of couplings was first recognized
in the theory of soliton equations \cite{Date:1982yeu} and appears
in various contexts of matrix models and related subjects.
For more details, see \cite{Okuyama:2021eju} and references therein.}
\begin{align}
t_k=\gamma_k+t(2k-1)!!(2\xi)^{-k-\frac{1}{2}}.
\label{eq:antitk}
\end{align}
This is valid as long as ${\rm Re}\,\xi>0$.
Here $t$ is the 't Hooft coupling in \eqref{eq:tHooft}.
Thus, \eqref{eq:TrAn} is evaluated as
\begin{align}
\begin{aligned}
\bra\Tr A^n\ket
 &\approx \bra\Tr A^n\ket^{g=0}\\
 &=\int_{E_0}^\infty dE\rho_0(E) A(E)^n,\\
 &\equiv Z_n,
\end{aligned}
\label{eq:Zn}
\end{align}
where $\rho_0(E)$ is now evaluated in the background \eqref{eq:antitk}.
In this background, \eqref{eq:f} becomes
\begin{align}
f(u=-v)\equiv f(v,t)=\rt{v}I_1(2\rt{v})+\frac{t}{\rt{2\xi+2v}},
\label{eq:f_anti}
\end{align}
where we have changed the variable as $v=-u$ for convenience and
$I_k(z)$ denotes the modified Bessel function of the first kind.
\eqref{eq:rho0} then becomes
\begin{align}
\rho_0(E)
 =\frac{1}{\sqrt{2}\pi\gs}\left(
  \int_{E_0}^E dv\frac{I_0(2\sqrt{v})}{\sqrt{E-v}}
 -\frac{t}{E+\xi}\sqrt{\frac{E-E_0}{2(E_0+\xi)}}\right).
\label{eq:rhoanti}
\end{align}
Note that in \cite{Okuyama:2021eju} we calculated $\rho_0(E)$
for JT gravity in the presence of $K$ FZZT branes.
The above $\rho_0(E)$ for anti-FZZT branes is essentially identical
to this except the sign of the 't~Hooft coupling $t$.
Note also that this expression of $\rho_0(E)$ is valid
as long as $t$ is not greater than the critical value $\tc$.
We will explain this in section~\ref{sec:thre}.

We emphasize that
we have treated anti-FZZT branes as dynamical objects.
More specifically, in \eqref{eq:TrAn} the color average is evaluated
in the presence of the determinant factor $\det(\xi+H)^{-K}$
and as a consequence
the deformed eigenvalue density \eqref{eq:rhoanti}
is used in \eqref{eq:Zn}.
This is the main difference from the approach of
\cite{Penington:2019kki},
which is based on the probe brane approximation
at genus-zero
\begin{equation}
\begin{aligned}
\bra\Tr A^n\ket_{\rm probe}\Big|_{g=0}&
 =\int dH e^{-\Tr V(H)} \Tr A^n\Big|_{g=0}\\
&= \int_0^\infty dE \rho_0^{\text{JT}}(E)A(E)^n,
\end{aligned} 
\label{eq:TrAn_probe}
\end{equation}
where the original JT gravity density of state
$\rho_0^{\text{JT}}(E)$ is given by
\begin{equation}
\begin{aligned}
 \rho_0^{\text{JT}}(E)=\frac{\sinh(2\rt{E})}{\rt{2}\pi\gs}.
\end{aligned} 
\label{eq:rho-JT}
\end{equation}
However, in \cite{Penington:2019kki}
the same 't Hooft limit as ours \eqref{eq:tHooft} is used.
As we argued in \cite{Okuyama:2021eju},
in this limit the back-reaction of
(anti-)FZZT branes cannot be ignored and the couplings $t_k$ are shifted
due to the insertion of branes.
As a consequence, the eigenvalue density is deformed from
$\rho_0^{\text{JT}}(E)$ in \eqref{eq:rho-JT} to
$\rho_0(E)$ in \eqref{eq:rhoanti}.
Thus we have to use the dynamical brane picture in this limit.

\subsection{Resolvent of reduced density matrix and entropy}

We saw in the last subsection that
the ensemble averages of $\varrho$ in the planar approximation
are expressed in terms of $Z_n$ in \eqref{eq:Zn}.
On the other hand, the general expression \eqref{eq:Tr-rho^n}
of $\bra\b{\Tr\varrho^n}\ket$ is rather complicated
as a function of $n$ and it is difficult
to apply the replica trick directly to \eqref{eq:Tr-rho^n}
to calculate the entropy.
Instead, as detailed in \cite{Penington:2019kki},
we can study the entropy using 
the resolvent $R(\la)$ for $\varrho$ in \eqref{eq:Rexp}.
By substituting \eqref{eq:Zn},
the Schwinger-Dyson equation \eqref{eq:SD-R} for $R(\la)$ becomes
\begin{equation}
\begin{aligned}
 \la R(\la)&=K+\sum_{n=1}^\infty
 \frac{R(\la)^n}{(KZ_1)^n}\int_{E_0}^\infty dE\rho_0(E)A(E)^n\\
&=K+\int_{E_0}^\infty dE\rho_0(E)\frac{w(E)R(\la)}{K-w(E)R(\la)},
\end{aligned} 
\label{eq:SD-int}
\end{equation}
where we have defined
\begin{equation}
\begin{aligned}
 w(E)=\frac{A(E)}{Z_1}.
\end{aligned} 
\end{equation}
Following \cite{Penington:2019kki}, we divide
the integral in \eqref{eq:SD-int} into two pieces
\begin{equation}
\begin{aligned}
 \la R(\la)&=K+\int_{E_0}^{E_K} dE\rho_0(E)\frac{w(E)R(\la)}{K-w(E)R(\la)}
+\int_{E_K}^{\infty} dE\rho_0(E)\frac{w(E)R(\la)}{K-w(E)R(\la)}\\
&\approx K+\int_{E_0}^{E_K} dE\rho_0(E)\frac{w(E)R(\la)}{K-w(E)R(\la)}
+\la_0 R(\la),
\end{aligned}
\label{eq:R-la0} 
\end{equation}
where $\la_0$ and $E_K$ are defined by
\begin{equation}
\begin{aligned}
 \la_0&=\frac{1}{K}\int_{E_K}^{\infty} dE\rho_0(E)w(E),\\
K&=\int_{E_0}^{E_K} dE\rho_0(E).
\end{aligned}
\label{eq:def-EKla0} 
\end{equation}
By rewriting \eqref{eq:R-la0} as
\begin{equation}
\begin{aligned}
R(\la)=\frac{K}{\la-\la_0}+\frac{1}{\la-\la_0}
 \int_{E_0}^{E_K} dE\rho_0(E)\frac{w(E)R(\la)}{K-w(E)R(\la)},
\end{aligned} 
\end{equation} 
we can solve $R(\la)$ by the iteration starting from $R(\la)=K/(\la-\la_0)$.
As discussed in \cite{Penington:2019kki}, the second order iteration gives
\begin{equation}
\begin{aligned}
 R(\la)&\approx \frac{K}{\la-\la_0}+\frac{1}{\la-\la_0}
 \int_{E_0}^{E_K} dE\rho_0(E)
 \frac{w(E)K(\la-\la_0)^{-1}}{K-w(E)K(\la-\la_0)^{-1}}\\
&=\frac{K}{\la-\la_0}+\frac{1}{\la-\la_0}\int_{E_0}^{E_K} dE
 \rho_0(E)\frac{w(E)}{\la-\la_0-w(E)}.
\end{aligned} 
\end{equation}
Using \eqref{eq:def-EKla0}, we find
\begin{equation}
\begin{aligned}
 R(\la)&=\frac{1}{\la-\la_0}\left[\int_{E_0}^{E_K} dE\rho_0(E)
+\int_{E_0}^{E_K} dE\rho_0(E)\frac{w(E)}{\la-\la_0-w(E)}\right]\\
&=\int_{E_0}^{E_K} dE\rho_0(E)\frac{1}{\la-\la_0-w(E)}.
\end{aligned} 
\end{equation} 
The eigenvalue density $D(\la)$ of the density matrix
$\varrho_{ij}$ is obtained from the discontinuity
of $R(\la)$
\begin{equation}
\begin{aligned}
 D(\la)=\frac{R(\la-\ri0)-R(\la+\ri0)}{2\pi\ri}=
\int_{E_0}^{E_K} dE\rho_0(E)\cob\bigl(\la-\la_0-w(E)\bigr).
\end{aligned} 
\label{eq:D-la}
\end{equation}
Finally, the von Neumann entropy is given by
\begin{equation}
\begin{aligned}
 S&=-\int d\la D(\la)\la\log\la\\
  &=-\int_{E_0}^{E_K} dE\rho_0(E)(\la_0+w(E))\log\bigl(\la_0+w(E)\bigr).
\end{aligned} 
\label{eq:S-PSSY}
\end{equation}
We will use this expression to study
the Page curve numerically in section~\ref{sec:page}.

In the rest of this section let us make several comments
on the above approximation.
We can check that $D(\la)$ in \eqref{eq:D-la} is normalized correctly
\begin{equation}
\begin{aligned}
 \Tr\varrho^0=\int d\la D(\la)\cdot1=\int_{E_0}^{E_K} dE\rho_0(E)=K,
\end{aligned} 
\end{equation}
where we have used \eqref{eq:def-EKla0}. We also find
\begin{equation}
\begin{aligned}
 \Tr\varrho&=\int d\la D(\la)\cdot\la=
\int_{E_0}^{E_K} dE\rho_0(E)(\la_0+w(E))\\
&=K\la_0+\int_{E_0}^{E_K} dE\rho_0(E)w(E)\\
&=\int_{E_K}^{\infty} dE\rho_0(E)w(E)+\int_{E_0}^{E_K} dE\rho_0(E)w(E)\\
&=\int_{E_0}^{\infty} dE\rho_0(E)w(E)\\
&=\frac{1}{Z_1}\int_{E_0}^{\infty} dE\rho_0(E)A(E)=\frac{1}{Z_1}Z_1=1,
\end{aligned} 
\label{eq:trrho-1}
\end{equation}
where we have used \eqref{eq:def-EKla0}.

$\la_0$ in \eqref{eq:def-EKla0} can be written as
\begin{equation}
\begin{aligned}
 \la_0&=\frac{1}{K}\int_{E_0}^{\infty}dE\rho_0(E)w(E)-
\frac{1}{K}\int_{E_0}^{E_K}dE\rho_0(E)w(E)\\
&=\frac{1}{K}-\b{w}_K,
\end{aligned} 
\end{equation}
where we have used \eqref{eq:trrho-1} and defined
\begin{equation}
\begin{aligned}
 \b{w}_K\equiv \frac{1}{K}\int_{E_0}^{E_K}dE\rho_0(E)w(E)
=\frac{\int_{E_0}^{E_K}dE\rho_0(E)w(E)}{\int_{E_0}^{E_K}dE\rho_0(E)}.
\end{aligned} 
\end{equation}
That is, $\b{w}_K$ is the average of $w(E)$ in the ``post Page''
subspace $E<E_K$.
Thus the resolvent is written as
\begin{equation}
\begin{aligned}
 R(\la)=\Tr\frac{1}{\la-\varrho}=\int_{E_0}^{E_K}dE\rho_0(E)
\frac{1}{\la-\la(E)},
\end{aligned} 
\end{equation}
where
\begin{equation}
\begin{aligned}
 \la(E)=\frac{1}{K}+w(E)-\b{w}_K.
\end{aligned} 
\label{eq:la-E}
\end{equation}
$\la(E)$ behaves as
\begin{equation}
\la(E)\approx \left\{
\begin{aligned}
 &\frac{1}{K}\quad &(K\ll \gs^{-1}),\\
&w(E)-\b{w}_K\quad &(K\gg \gs^{-1}).
\end{aligned} \right.
\end{equation}
This corresponds to Figure~6 in \cite{Penington:2019kki}.
Note that the density matrix $\varrho_{ij}$
is originally a matrix in the flavor space, but after taking the average
the spectrum $\la(E)$ of $\varrho_{ij}$ is effectively written
in terms of energy eigenvalues in the ``color'' space.
We have to project quantities onto the ``post Page'' subspace
$E<E_K$ to ensure that the number of total state is $K=\int_{E<E_K}
dE\rho_0(E)$.

\section{Phase transition}\label{sec:phase}

An interesting feature of the anti-FZZT brane background
in JT gravity is that
the system exhibits a phase transition as the 't Hooft coupling $t$
varies. In this section we discuss this phase transition
and study the critical behavior of the entropy.

\subsection{Threshold energy}\label{sec:thre}

\begin{figure}[t]
\centering
\[
\includegraphics[width=0.7\linewidth]{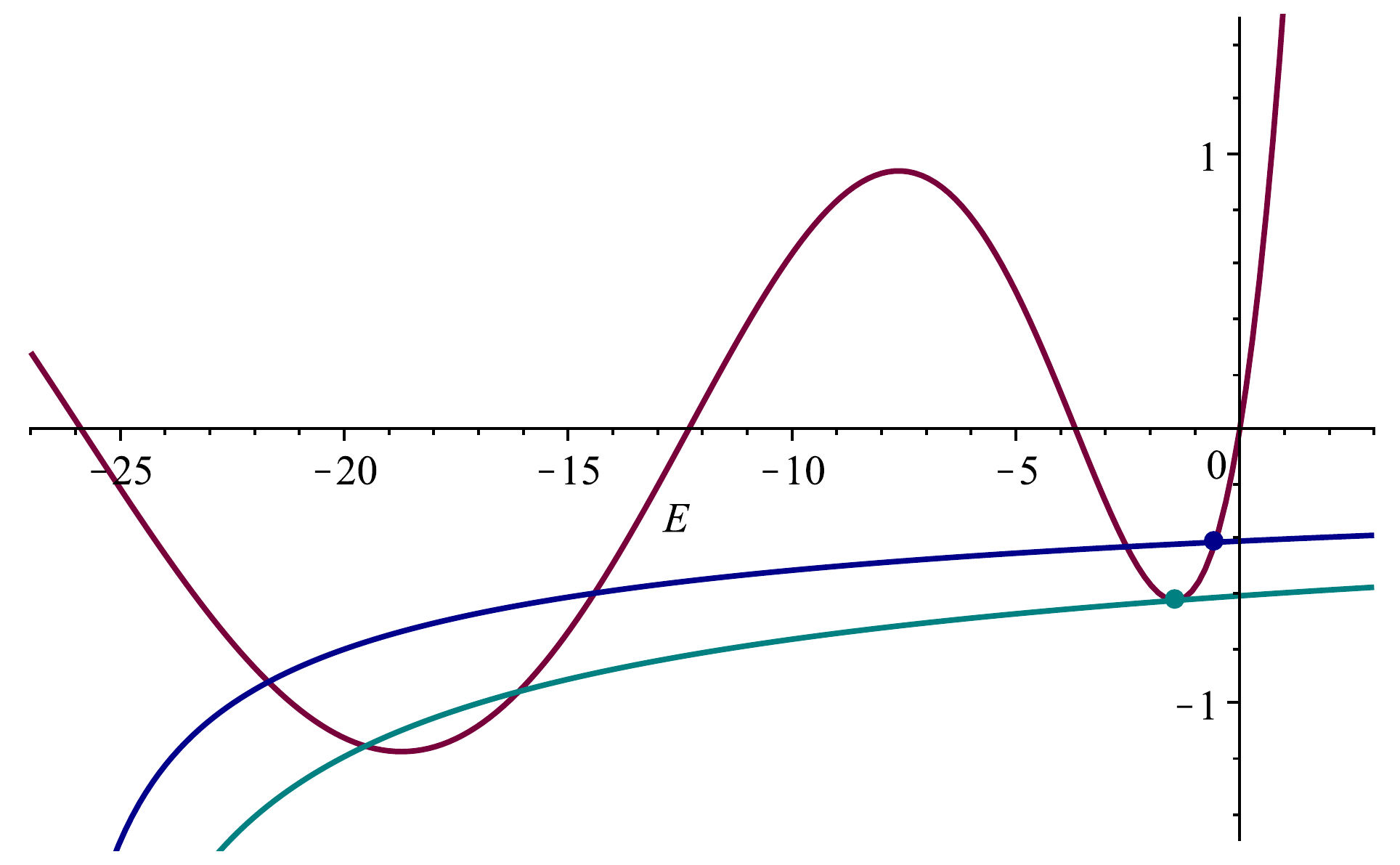}
\]
\caption{
The threshold energy $E_0$ is determined by
the equation \eqref{eq:threanti}.
The red curve represents the graph of $\sqrt{E}I_1(2\sqrt{E})$
while the blue and green curves are the graphs of $-t/\sqrt{2(E+\xi)}$
with $t=3$ and $t=\tc\approx 4.46$
respectively, where we set $\xi=27$.
As we see in this example,
\eqref{eq:threanti} could have multiple real solutions.
The threshold energy $E_0$ is determined as the largest real solution,
as indicated by a dot.
The horizontal location of the green dot gives the critical value $\Ec$.}
\label{fig:streq}
\end{figure}

Let us first clarify the definition of the threshold energy $E_0$.
In the last section we saw that $E_0$ is determined by
the threshold energy condition \eqref{eq:E0cond}.
For JT gravity with anti-FZZT branes,
$f$ is given by \eqref{eq:f_anti}
and the condition \eqref{eq:E0cond} is written explicitly as
\begin{align}
\sqrt{E_0}I_1(2\sqrt{E_0})=-\frac{t}{\sqrt{2(E_0+\xi)}}.
\label{eq:threanti}
\end{align}
The threshold energy $E_0$ is determined as a real solution of 
this equation.
Here, $t>0$ by definition and we take $\xi>0$ in order for 
the shift of the couplings \eqref{eq:antitk} to be valid.
We show the plots of both sides of the equation \eqref{eq:threanti}
in Figure~\ref{fig:streq}.
We see that $\sqrt{E}I_1(2\sqrt{E})$ oscillates for $E<0$
and increases monotonically for $E>0$ starting from the origin,
while $-t/\sqrt{2(E+\xi)}$ is always negative.
Therefore $E_0$ has to be negative if it exists.
However,
the number of real solutions of \eqref{eq:threanti} varies
depending on the values of $t$ and $\xi$. In particular,
\eqref{eq:threanti} has no real solution when $t$ is very large,
whereas it has multiple real solutions when $\xi$ is large
and $t$ is small.\footnote{See also \cite{Johnson:2020lns}
for a similar problem in JT gravity with conical defects.}
On the other hand, one can easily see that
\eqref{eq:threanti} always has at least one real solution
for sufficiently small $t$.
We define $E_0$ as the largest real solution of \eqref{eq:threanti}
(i.e.~the solution with the smallest absolute value),
so that it is continuously deformed from $E_0=0$
for the original JT gravity case $t=0$.

As we see from Figure~\ref{fig:streq},
$|E_0|$ is small for small $t$.
If we increase $t$, $|E_0|$ also increases.
Then there exists a critical point $t=\tc$ beyond which $E_0$
no longer continues as a real solution.
Thus, we expect a phase transition.
This transition is qualitatively very similar to the one
discussed in \cite{Gao:2021uro} in the case of EOW branes.
If one continuously increases $t$ beyond the critical point,
$E_0$ and the second largest root turn into a pair of complex roots.
It is therefore very likely that
the saddle point of the matrix integral is described by
an eigenvalue density with ``Y'' shaped support,
similar to the one studied in \cite{Gao:2021uro}.
It would be interesting to study the model
in this ``Y'' shaped phase further.
In this paper we view this phase transition as
the end of the black hole evaporation
and focus on the physics before the phase transition.

\subsection{Behavior of threshold energy near $t=\tc$}

At the critical value $t=\tc$, the equation \eqref{eq:threanti}
has a double root $\Ec$.
Thus $\tc$ is determined by the condition
\begin{equation}
\begin{aligned}
 f(\Ec,\tc)=0,\quad \del_{v}f(\Ec,\tc)=0
\end{aligned} 
\label{eq:double-root}
\end{equation}
with $f(v,t)$ given in \eqref{eq:f_anti}.
Expanding the equation $f(E_0,t)=0$ around $(v,t)=(\Ec,\tc)$, we find
\begin{equation}
\begin{aligned}
 0&=f(\Ec,\tc)+\del_{v}f(\Ec,\tc)(E_0-\Ec)
   +\hf \del_{v}^2f(\Ec,\tc)(E_0-\Ec)^2\\
&\hspace{1em}+\del_tf(\Ec,\tc)(t-\tc)+\cdots.
\end{aligned} 
\end{equation}
The first two terms vanish due to \eqref{eq:double-root}.
Thus, near $t=\tc$ we find
\begin{equation}
\begin{aligned}
 E_0-\Ec\approx C \rt{\tc-t}\quad(t<\tc),
\end{aligned} 
\label{eq:E0behavior}
\end{equation}
where $C$ is given by
\begin{equation}
\begin{aligned}
 C=\rt{\frac{2\del_tf(\Ec,\tc)}{\del_{v}^2f(\Ec,\tc)}}.
\end{aligned} 
\end{equation}

\subsection{Effective zero-temperature entropy
 and von Neumann entropy near $t=\tc$}

\begin{figure}[t]
\centering
\includegraphics[width=0.7\linewidth]{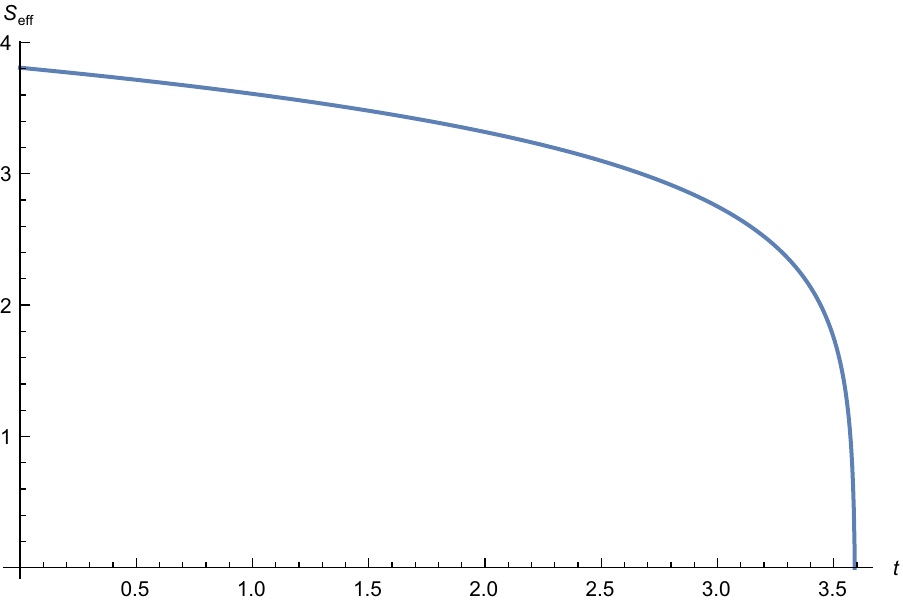}
  \caption{
Plot 
of the effective entropy $S_{\text{eff}}$.
We set $\xi=18,\,\gs=1/100$ in this figure.
}
  \label{fig:Seff}
\end{figure}

In \cite{Gao:2021uro},
the effective zero-temperature entropy $S_{\text{eff}}$ 
was introduced. It is defined 
by the behavior of $\rho_0(E)$ near $E=E_0$:
\begin{equation}
\begin{aligned}
 \rho_0(E)\sim e^{S_{\text{eff}}}\rt{E-E_0}.
\end{aligned} 
\label{eq:rho0behavior}
\end{equation}
From \eqref{eq:rho0} we find
\begin{equation}
\begin{aligned}
 e^{S_{\text{eff}}}&=\frac{\rt{2}}{\pi \gs}\partial_v f(E_0,t)\\
&=\frac{\rt{2}}{\pi \gs}\left[I_0(2\rt{E_0})
  -\frac{t}{(2\xi+2E_0)^{\frac{3}{2}}}\right].
\end{aligned} 
\end{equation}
In Figure~\ref{fig:Seff}, we show the plot of $S_{\text{eff}}$.
We see that $S_{\text{eff}}$ is a monotonically
decreasing function of $t$.
Near $t=\tc$, using \eqref{eq:double-root} and \eqref{eq:E0behavior}
we find 
\begin{align}
\begin{aligned}
e^{S_{\text{eff}}}
&\sim\partial_v f(E_0,t)\\
&=\partial_v f(\Ec,\tc)+\partial_v^2 f(\Ec,\tc)(E_0-\Ec)
 +\partial_t\partial_v f(\Ec,\tc)(t-\tc)+\cdots\\
&\sim \rt{\tc-t}.
\end{aligned}
\label{eq:Seffbehavior}
\end{align}

Using the above results, we can evaluate the
critical behavior of the von Neumann entropy \eqref{eq:S-PSSY}.
It turns out that the critical behavior of the von Neumann entropy 
$S(t)$ in \eqref{eq:S-PSSY}
is determined by that of the eigenvalue density
near $E=E_0$
\begin{equation}
\begin{aligned}
 \rho_0(E)\sim e^{S_{\text{eff}}}\rt{E-E_0}\sim\rt{\tc-t}\rt{E-E_0}.
\end{aligned} 
\label{eq:rho-crit}
\end{equation}
One can show that the contribution of
the $E$-integral \eqref{eq:S-PSSY} away from the edge $E=E_0$
is finite at $t=\tc$. Subtracting this finite contribution
and using \eqref{eq:rho-crit} near $E=E_0$ in \eqref{eq:S-PSSY},
we find
\begin{align}
S(t)-S(\tc)\sim\sqrt{\tc-t}.
\label{eq:Sbehavior}
\end{align}
In the next section we will confirm this behavior numerically.

\section{Numerical study of Page curve}\label{sec:page}

In section~\ref{sec:setup} we saw how to calculate the entropy.
In this section we will numerically study the Page curve,
i.e.~the time evolution of the entropy.
In Page's original calculation $\log K$ is regarded as ``time''
\cite{Page:1993df}.
Since we take the 't Hooft limit \eqref{eq:tHooft},
we will regard $\log t$ as ``time.''
We will plot the entropy as a function of $t$
rather than $\log t$, which is convenient for seeing
the critical behavior discussed in section~\ref{sec:phase}.

\subsection{Von Neumann entropy}

\begin{figure}[t]
\centering
\includegraphics[width=0.7\linewidth]{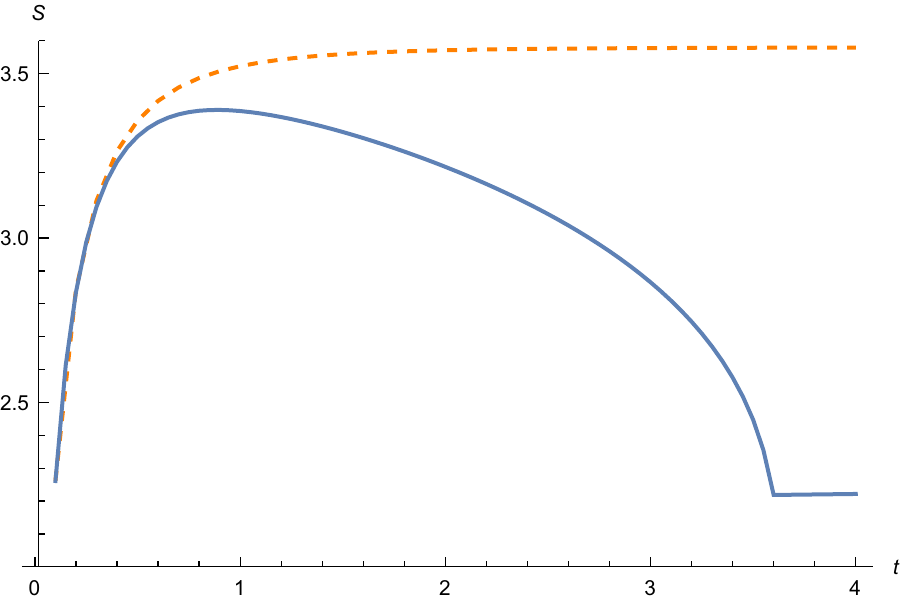}
  \caption{
Plot 
of the von Neumann entropy $S$ in \eqref{eq:S-PSSY} 
as a function of $t= \gs K$. 
We set $\xi=18,\,\bt=4,\,\gs=1/100$ in this figure.
The solid blue curve is the dynamical brane case
while the dashed orange curve is the probe brane case.
}
  \label{fig:S-antiFZZT}
\end{figure}

We consider the von Neumann entropy \eqref{eq:S-PSSY}
in JT gravity in the presence of $K$ anti-FZZT branes.
As discussed in section~\ref{sec:setup}
we compute the entropy for dynamical branes, but it is interesting to
compute it in the probe brane approximation as well
for the sake of comparison.
In the probe brane approximation, we have $E_0=0$ and
the eigenvalue density is given by
$\rho_0^{\text{JT}}(E)$ in \eqref{eq:rho-JT}.
We show the plot of the von Neumann entropy $S$
in Figure~\ref{fig:S-antiFZZT}.
We see that the entropy for the dynamical brane 
(solid blue curve) starts to decrease
relative to the probe brane case (dashed orange curve). 
This is very similar to the Page curve of an evaporating black hole.
As we saw in the last section, we observe that
the entropy exhibits the critical behavior~\eqref{eq:Sbehavior}.

\subsection{R\'{e}nyi entropy}

It is also interesting to consider the R\'{e}nyi entropy.
The $n$th R\'{e}nyi entropy $S_n$ is defined by
\begin{equation}
\begin{aligned}
\bra\b{\Tr\varrho^n}\ket=e^{-(n-1)S_n}.
\end{aligned} 
\label{eq:Sn}
\end{equation}
For large $K$, $\bra\b{\Tr\varrho^n}\ket$ is dominated
by the last term of \eqref{eq:Tr-rho^n}.
Thus as $t$ grows, $S_n$ approaches
\begin{equation}
\begin{aligned}
 \til{S}_n:=-\frac{1}{n-1}\log\frac{Z_n}{(Z_1)^n}.
\end{aligned} 
\label{eq:tSn}
\end{equation}

\begin{figure}[t]
\centering
\includegraphics[width=0.7\linewidth]{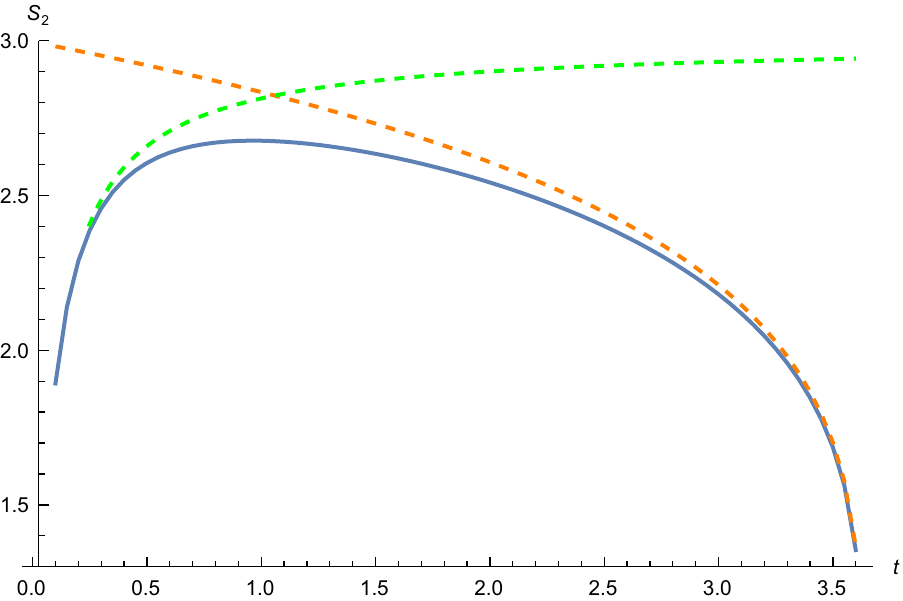}
  \caption{
Plot 
of the second R\'{e}nyi entropy $S_2$ in \eqref{eq:S2}
as a function of $t= \gs K$. 
We set $\xi=18,\,\bt=4,\,\gs=1/100$ in this figure.
The solid blue curve is $S_2$ in \eqref{eq:S2}.
The dashed orange curve represents $\til{S}_2$ in \eqref{eq:tSn}
while the green dashed curve is $S_2$
without taking account of the back-reaction.
}
  \label{fig:S2-antiFZZT}
\end{figure}

Let us first consider the second R\'{e}nyi entropy
\begin{equation}
\begin{aligned}
 S_2=-\log\left(\frac{\gs}{t}+\frac{Z_2}{Z_1^2}\right).
\end{aligned} 
\label{eq:S2}
\end{equation}
In Figure~\ref{fig:S2-antiFZZT}, we show the plot of $S_2$
as a function of $t$.
We can see that $S_2$ first increases and then decreases.
Near $t=\tc$, $S_2$ exhibits a critical behavior 
\begin{equation}
\begin{aligned}
 S_2(t)-S_2(\tc)\sim \sqrt{\tc-t}.
\end{aligned} 
\label{eq:S-gamma}
\end{equation}
This behavior can be derived in the same way
as in the case of the von Neumann entropy.

We can see that the plot of $S_2$ 
has a similar behavior with the Page curve of Hawking radiation
from an evaporating black hole
(see e.g.~Figure~7 in \cite{Almheiri:2020cfm}).
We regard $\til{S}_2$ as an analogue of
the thermodynamic entropy $S_{\text{BH}}$ of an evaporating black hole.
From Figure~\ref{fig:S2-antiFZZT}, we see that $\til{S}_2$ 
is a monotonically decreasing function of $t$
(represented by the dashed orange curve).

\begin{figure}[t]
\centering
\includegraphics[width=0.7\linewidth]{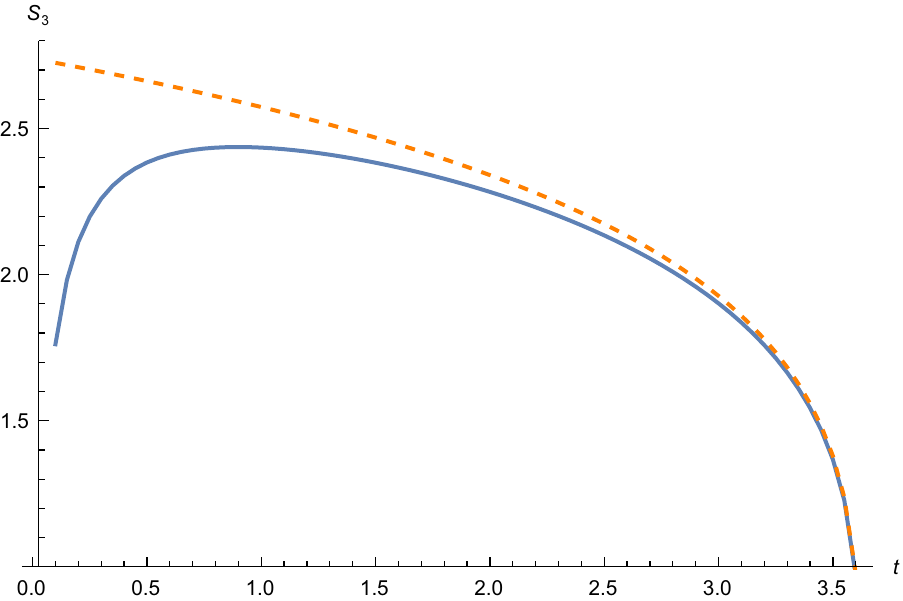}
  \caption{
Plot 
of the third R\'{e}nyi entropy $S_3$ in \eqref{eq:S3}
as a function of $t= \gs K$. 
We set $\xi=18,\,\bt=4,\,\gs=1/100$ in this figure.
The solid blue curve is $S_3$ in \eqref{eq:S3}.
The dashed orange curve represents $\til{S}_3$ in \eqref{eq:tilS3}.
}
  \label{fig:S3-antiFZZT}
\end{figure}

Let us next consider the third R\'{e}nyi entropy
\begin{equation}
\begin{aligned}
 e^{-2S_3}=\frac{1}{K^2}+\frac{3Z_2}{KZ_1^2}+\frac{Z_3}{Z_1^3}.
\end{aligned} 
\label{eq:S3}
\end{equation}
As $t$ grows, this approaches
\begin{equation}
\begin{aligned}
 e^{-2\til{S}_3}=\frac{Z_3}{Z_1^3}.
\end{aligned} 
\label{eq:tilS3}
\end{equation}
In Figure~\ref{fig:S3-antiFZZT}
we show the plot of $S_3$ for anti-FZZT branes.
We see that this is qualitatively very similar to the $S_2$ case.
In particular, $S_3$ is bounded from above by $\tS_3$, which
monotonically decreases.
From Figure~\ref{fig:S2-antiFZZT} and \ref{fig:S3-antiFZZT},
it is natural to regard $\til{S}_n$ in \eqref{eq:tSn}
as an analogue of the thermodynamic entropy $S_{\text{BH}}$
of black hole, since $S_{\text{BH}}$ decreases monotonically
during the evaporation process as well
\begin{equation}
\begin{aligned}
 \til{S}_n
 ~\leftrightarrow~S_{\text{BH}}.
\end{aligned} 
\label{eq:Sn-SBH}
\end{equation}

Based on the above numerical results,
we conjecture that $\til{S}_n$ defined in \eqref{eq:tSn}
is a monotonically decreasing function of $t$.
More specifically, we conjecture that
\begin{align}
\partial_t \tS_n<0 \quad\mbox{for}\quad
n > 1,\quad 0<t<\tc.
\label{eq:tSconj}
\end{align}
We will study this monotonic behavior in the next section.

\section{\mathversion{bold}Monotonicity of $\til{S}_n$}\label{sec:mono}

In this section we study the monotonically decreasing behavior of $\tS_n$.
We will prove our conjecture \eqref{eq:tSconj}
in the large $\xi$ limit.
We will also discuss how to understand
intuitively this monotonically decreasing behavior.

\subsection{Leading-order eigenvalue density and its derivative}

In this subsection let us derive some useful formulas
about the leading-order eigenvalue density $\rho_0(E)$ in \eqref{eq:rho0}
for Witten-Kontsevich gravity with general couplings $\{t_k\}$,
which we will use in the next subsection.

To study general Witten-Kontsevich gravity
it is convenient to introduce the Itzykson-Zuber variables
\cite{Itzykson:1992ya}
\begin{align}
\cI_n(u)\equiv \cI_n(u,\{t_k\}) =\sum_{m=0}^\infty t_{n+m}\frac{u^m}{m!}
\quad (n\ge 0).
\label{eq:defIn}
\end{align}
In terms of $\cI_n$,
$f(u)$ in \eqref{eq:f} is written as
\begin{align}
f(u)&=u-\cI_0(u)
\label{eq:finI}
\end{align}
and the leading-order density $\rho_0(E)$ in \eqref{eq:rho0}
becomes
\begin{equation}
\begin{aligned}
 \rho_0(E)
&=\frac{1}{\rt{2}\pi \gs}\int_{E_0}^E dv\frac{1-\cI_1(-v)}{\rt{E-v}}.
\end{aligned} 
\label{eq:rho0-sum}
\end{equation}

Let us now consider the anti-FZZT brane background \eqref{eq:antitk}.
We are interested in how the entropy evolves
as the 't Hooft coupling $t$ in \eqref{eq:tHooft} grows.
Note that $t$ is implicitly related to $E_0$ by
the string equation \eqref{eq:E0cond},
from which one finds
\begin{equation}
\begin{aligned}
0&=\del_tf(-E_0)\\
 &=(\del_tE_0)(\cI_1(-E_0)-1)-\del_t\cI_0(u)\Big|_{u=-E_0}\\
 &=(\del_tE_0)(\cI_1(-E_0)-1)-\frac{1}{\rt{2\xi+2E_0}}.
\end{aligned}
\label{eq:deltE0}
\end{equation}
By using this relation, the $t$-derivative of $\rho_0(E)$
is calculated as
\begin{equation}
\begin{aligned}
 \del_t\rho_0(E)
&=(\del_tE_0)\partial_{E_0}\rho_0(E)
 -\frac{1}{\rt{2}\pi \gs}\int_{E_0}^E dv
  \frac{\partial_t\cI_1(-v)}{\rt{E-v}}\\
&=\frac{\del_tE_0}{\sqrt{2}\pi \gs}\frac{\cI_1(-E_0)-1}{\rt{E-E_0}}
 -\frac{1}{\sqrt{2}\pi\gs}\int_{E_0}^E dv(E-v)^{-\frac{1}{2}}
  (2\xi+2v)^{-\frac{3}{2}}\\
&=\frac{1}{2\pi\gs\sqrt{(E-E_0)(\xi+E_0)}}
 -\frac{1}{2\pi\gs(E+\xi)}\sqrt{\frac{E-E_0}{\xi+E_0}}\\
&=\frac{1}{2\pi \gs(E+\xi)}\sqrt{\frac{\xi+E_0}{E-E_0}}.
\end{aligned} 
\label{eq:drho0}
\end{equation}
Note that the background \eqref{eq:antitk} is written for JT gravity,
but we have never used the specific values
of $\gamma_k$ in the above derivation.
Therefore, \eqref{eq:drho0} is in fact valid for the anti-FZZT brane
background of other gravities as well.

\subsection{Proof in the large $\xi$ limit}

In the last section we conjectured that $\tS_n$ defined in \eqref{eq:tSn}
is a monotonically decreasing function of $t$.
In this subsection let us prove this conjecture \eqref{eq:tSconj}
at large $\xi$.
From the expression \eqref{eq:tSn},
we see that \eqref{eq:tSconj} is equivalent to
\begin{align}
\frac{\partial_t Z_n}{nZ_n}>\frac{\partial_t Z_1}{Z_1}.
\label{eq:tSconj2}
\end{align}

By using the property $\rho_0(E_0)=0$ and the expression \eqref{eq:drho0},
the $t$-derivative of $Z_n$ in \eqref{eq:Zn}
is calculated as
\begin{align}
\begin{aligned}
\partial_t Z_n
 &=\int_{E_0}^\infty dE\partial_t\rho_0(E)A(E)^n\\
 &=\int_{E_0}^\infty dE
  \frac{1}{2\pi\gs}\sqrt{\frac{\xi+E_0}{E-E_0}}
  \frac{e^{-n\beta E}}{(E+\xi)^{n+1}}\\
 &=\frac{\sqrt{\xi+E_0}}{2\pi\gs}e^{-n\beta E_0}
   \int_0^\infty d\tE \tE^{-\frac{1}{2}}
   (\tE+E_0+\xi)^{-n-1}e^{-n\beta\tE},
\end{aligned}
\label{eq:dtZn}
\end{align}
where we have set $\tE=E-E_0$.
For large $\xi$, \eqref{eq:dtZn} is evaluated as
\begin{align}
\begin{aligned}
\partial_t Z_n
&=\frac{e^{-n\beta E_0}}{2\sqrt{n\pi\beta}\gs}\xi^{-n-\frac{1}{2}}
  +{\cal O}\left(\xi^{-n-\frac{3}{2}}\right).
\end{aligned}
\label{eq:tZn_laxi}
\end{align}

On the other hand, by plugging \eqref{eq:rhoanti}
into \eqref{eq:Zn}, $Z_n$ is written as
\begin{align}
\begin{aligned}
Z_n&=\int_{E_0}^\infty\rho_0(E)A(E)^n\\
 &=
  \frac{1}{\sqrt{2}\pi\gs}\int_{E_0}^\infty dv I_0(2\sqrt{v})
  \int_v^\infty dE(E-v)^{-\frac{1}{2}}(E+\xi)^{-n}e^{-n\beta E}\\
&\hspace{1em}
 -\frac{t}{2\pi\gs(E_0+\xi)^{\frac{1}{2}}}
  \int_{E_0}^\infty dE(E-E_0)^{\frac{1}{2}}(E+\xi)^{-n-1}e^{-n\beta E}.
\end{aligned}
\label{eq:Znanti}
\end{align}
In the same way as in \eqref{eq:dtZn}--\eqref{eq:tZn_laxi},
the above integrals at large $\xi$ are evaluated as
\begin{align}
\begin{aligned}
Z_n
&=\frac{1}{\sqrt{2n\pi\beta}\gs}
  \int_{E_0}^\infty dv I_0(2\sqrt{v})(v+\xi)^{-n}e^{-n\beta v}
  +{\cal O}\left(\xi^{-n-1}\right)\\
&\hspace{1em}
 -\frac{te^{-n\beta E_0}}{4\sqrt{\pi}\gs(n\beta)^{\frac{3}{2}}}
  \xi^{-n-\frac{3}{2}}
  +{\cal O}\left(\xi^{-n-\frac{5}{2}}\right)\\
&=\frac{1}{\sqrt{2n\pi\beta}\gs\xi^n}
  \int_{E_0}^\infty dv I_0(2\sqrt{v})e^{-n\beta v}
  +{\cal O}\left(\xi^{-n-1}\right).
\end{aligned}
\label{eq:Zn_laxi}
\end{align}
Here we see that in the leading-order of the large-$\xi$ approximation
the first integral in \eqref{eq:Znanti}
is dominant and the second integral does not contribute.
Indeed, from \eqref{eq:Zn_laxi} one can reproduce
\eqref{eq:tZn_laxi} by using the relation
$\partial_t E_0= -\left(\sqrt{2\xi}I_0(2\sqrt{E_0})\right)^{-1}
+{\cal O}(\xi^{-3/2})$,
which follows from \eqref{eq:threanti} or \eqref{eq:deltE0}.
Thus we obtain
\begin{align}
\begin{aligned}
\frac{\partial_t Z_n}{n Z_n}
&=\frac{e^{-n\beta E_0}}
 {n\sqrt{2\xi}\int_{E_0}^\infty dv I_0(2\sqrt{v})e^{-n\beta v}}
 +{\cal O}\left(\xi^{-\frac{3}{2}}\right).
\end{aligned}
\label{eq:dZnoverZn}
\end{align}

To prove \eqref{eq:tSconj2} at large $\xi$,
it is sufficient to show that
\eqref{eq:dZnoverZn} monotonically increases as $n$ grows:
\begin{align}
\partial_n\frac{\partial_t Z_n}{n Z_n}>0.
\label{eq:lxieneq0}
\end{align}
Since ${\partial_t Z_n}/{n Z_n}>0$,\footnote{This
follows from \eqref{eq:dZnoverZn}
with $I_0(2\sqrt{v})=\partial_v[\sqrt{v}I_1(2\sqrt{v})]>0$
for $v>\Ec$, as we can see
from Figure~\ref{fig:streq}.\label{footnote:pos}}
this is equivalent to showing that
\begin{align}
\partial_n\log\frac{\partial_t Z_n}{n Z_n}>0.
\label{eq:lxieneq1}
\end{align}
The l.h.s.~of \eqref{eq:lxieneq1} is rewritten as
\begin{align}
\begin{aligned}
\partial_n\log\frac{\partial_t Z_n}{n Z_n}
 &=-\beta E_0-\frac{1}{n}
  +\frac{\int_{E_0}^\infty dv I_0(2\sqrt{v})\beta v e^{-n\beta v}}
        {\int_{E_0}^\infty dv I_0(2\sqrt{v})e^{-n\beta v}}\\
 &=\frac{\int_{E_0}^\infty dv I_0(2\sqrt{v})
         \left[n\beta(v-E_0)-1\right] e^{-n\beta v}}
        {n\int_{E_0}^\infty dv I_0(2\sqrt{v})e^{-n\beta v}}\\
 &=\frac{\int_0^\infty d\tv I_0(2\sqrt{\tv+E_0})
         \left(n\beta\tv-1\right) e^{-n\beta\tv}}
        {ne^{n\beta E_0}\int_{E_0}^\infty dv I_0(2\sqrt{v})e^{-n\beta v}},
\end{aligned}
\label{eq:lxieneq2}
\end{align}
where we have set $\tv=v-E_0$.
The denominator of the last expression in \eqref{eq:lxieneq2} is positive
(see footnote~\ref{footnote:pos}).
By renaming $\tv$ as $v$, the numerator is evaluated as
\begin{align}
\begin{aligned}
&\int_0^\infty dv I_0(2\sqrt{v+E_0})
         n\beta v e^{-n\beta v}
 -\int_0^\infty dv I_0(2\sqrt{v+E_0})e^{-n\beta v}\\
&=-I_0(2\sqrt{v+E_0})v e^{-n\beta v}\Big|_{0}^\infty
 +\int_0^\infty dv\left(I_0(2\sqrt{v+E_0})
     +\frac{I_1(2\sqrt{v+E_0})}{\sqrt{v+E_0}}v\right) e^{-n\beta v}\\
&\hspace{1em}
-\int_0^\infty dv I_0(2\sqrt{v+E_0})e^{-n\beta v}\\
&=\int_0^\infty dv
     \frac{I_1(2\sqrt{v+E_0})}{\sqrt{v+E_0}}v e^{-n\beta v}.
\end{aligned}
\label{eq:lxieneq3}
\end{align}
Since the integrand is positive
for any $E_0$ satisfying $\Ec<E_0<0$,\footnote{This is easily seen
from the graph of
$\sqrt{E}I_1(2\sqrt{E})=\left(I_1(2\sqrt{E})/\sqrt{E}\right)\times E$
in Figure~\ref{fig:streq}.}
\eqref{eq:lxieneq3} is positive.
Thus we have proved \eqref{eq:lxieneq1}.
Hence \eqref{eq:tSconj} has been proved at large $\xi$.

\subsection{Intuitive understanding of
 the decreasing behavior of $\til{S}_n$}

Beyond the large $\xi$ approximation,
it does not seem easy to find a simple analytic proof of
the monotonically decreasing behavior of $\tS_n$.
Alternatively, in this subsection we will explain how to understand
intuitively the monotonically decreasing behavior of $\tS_n$.
In contrast to the proof in the last subsection, the idea we will
describe does not depend on the details of the JT gravity background
and thus it can be generalized to the other gravity cases as well.

The replica index $n$ is sometimes identified as an analogue of the
inverse temperature (see e.g.~\cite{Nakaguchi:2016zqi}).
Here we will pursue this analogy.
To do this, let us consider the change of variable from $E$ to $\cE$
given by\footnote{\eqref{eq:cE-AE} is rewritten as
\begin{equation}
\begin{aligned}
 \bt(\xi+E)e^{\bt(\xi+E)}=\bt e^{\cE+\bt\xi}.
\end{aligned} 
\end{equation}
This is solved by the Lambert function $W(z)e^{W(z)}=z$ as
\begin{equation}
\begin{aligned}
 \bt(\xi+E)=W\Bigl(\bt e^{\cE+\bt\xi}\Bigr).
\end{aligned} 
\end{equation}
}
\begin{equation}
\begin{aligned}
 e^{-\cE}=A(E)=\frac{e^{-\bt E}}{\xi+E}.
\end{aligned} 
\label{eq:cE-AE}
\end{equation}
Then we find
\begin{equation}
\begin{aligned}
 Z_n&=\int_{E_0}^\infty dE\rho_0(E)A(E)^n=\int_{\cE_0}^\infty d\cE
D(\cE)e^{-n\cE},
\end{aligned} 
\label{eq:Zn-DE}
\end{equation}
where $\cE_0=-\log A(E_0)$ and 
\begin{equation}
\begin{aligned}
 D(\cE)=\left(\frac{\del \cE}{\del E}\right)^{-1}\rho_0(E)
 =\frac{\xi+E}{\bt(\xi+E)+1}\rho_0(E).
\end{aligned} 
\label{eq:D-rho}
\end{equation}
$Z_n$ in \eqref{eq:Zn-DE} takes the form of the canonical partition function
with inverse temperature $n$ and density of states $D(\cE)$.
In this picture the ``thermodynamic entropy'' is expressed as
\cite{Dong:2016fnf,Nakaguchi:2016zqi}
\begin{equation}
\begin{aligned}
 S_{\text{therm}}=(1-n\del_n)\log Z_n.
\end{aligned} 
\end{equation}
On the other hand, as we saw in the last subsection
$\del_t\til{S}_n<0$ is equivalent to \eqref{eq:lxieneq0},
which is written as
\begin{equation}
\begin{aligned}
-\del_n \frac{\del_tZ_n}{nZ_n}&=
\frac{1}{n^2}\del_t S_{\text{therm}}<0.
\end{aligned} 
\end{equation}
Therefore, the monotonically decreasing behavior of $\tS_n$
is interpreted as
that of the thermodynamic entropy $S_{\text{therm}}$.

Let us list some useful properties of $S_{\text{therm}}$:
\begin{enumerate}
 \item The threshold energy $\cE_0$ does not contribute to $S_{\text{therm}}$:
If we define $\til{Z}_n$ by
\begin{equation}
\begin{aligned}
 Z_n=e^{-n\cE_0}\til{Z}_n,\quad
\til{Z}_n=\int_0^\infty d\cE D(\cE+\cE_0)e^{-n\cE},
\end{aligned} 
\end{equation}
then we find
\begin{equation}
\begin{aligned}
 S_{\text{therm}}=(1-n\del_n)(-n\cE_0+\log \til{Z}_n)=(1-n\del_n)
\log \til{Z}_n.
\end{aligned} 
\end{equation}
\item The overall scale of $\til{Z}_n$ does contribute to $S_{\text{therm}}$:
If we define
\begin{equation}
\begin{aligned} 
\til{Z}_n=e^{S(t)}Z_n',
\end{aligned} 
\end{equation}
where $S(t)$ is $n$-independent, then we find
\begin{equation}
\begin{aligned}
 S_{\text{therm}}=(1-n\del_n)(S(t)+\log Z_n')=S(t)+(1-n\del_n)\log Z_n'.
\end{aligned} 
\end{equation}
If we further assume that $Z_n'$ is $t$-independent,
then the monotonically decreasing behavior of $S(t)$
implies that of $S_{\text{therm}}$. 

\item $S_{\text{therm}}$ is written as
\begin{equation}
\begin{aligned}
 S_{\text{therm}}=\log\til{Z}_n+n\bra \cE\ket,
\end{aligned} 
\label{eq:S-tZrel}
\end{equation}
where $\bra \cE\ket$ is given by
\begin{equation}
\begin{aligned}
 \bra \cE\ket=-\del_n\log\til{Z}_n=\frac{1}{\til{Z}_n}\int_0^\infty d\cE
D(\cE+\cE_0)e^{-n\cE}\cE.
\end{aligned} 
\end{equation}
\end{enumerate}

\begin{figure}[t]
\centering
\subcaptionbox{$\rho_0(E)$ vs $E$\label{sfig:rho0}}{\includegraphics
[width=0.45\linewidth]{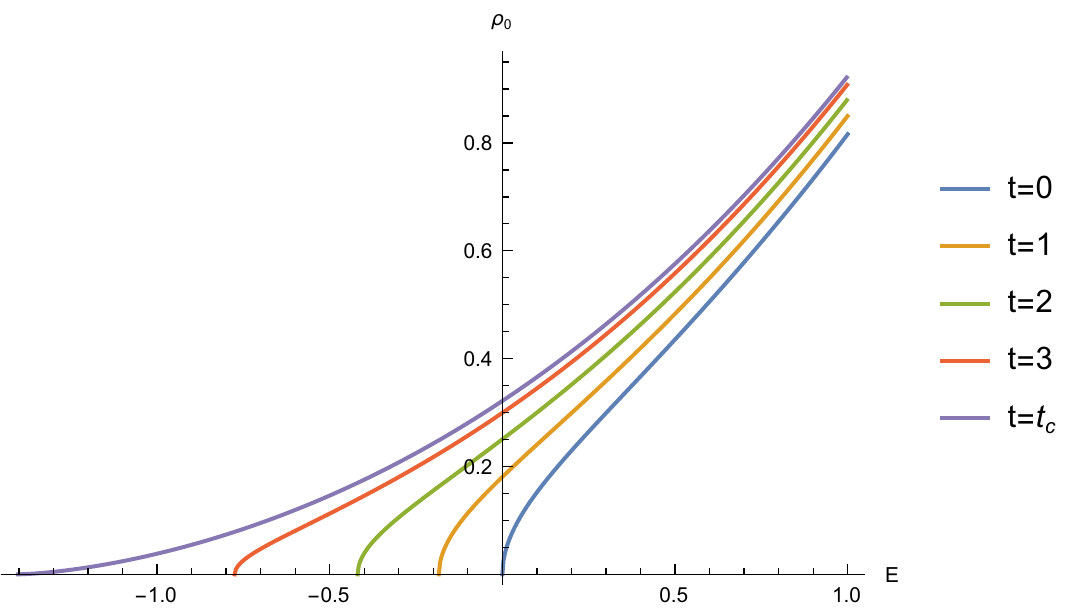}}
\hskip5mm
\subcaptionbox{$\rho_0(E)$ vs $E-E_0$\label{sfig:rho0-shift}}{\includegraphics
[width=0.45\linewidth]{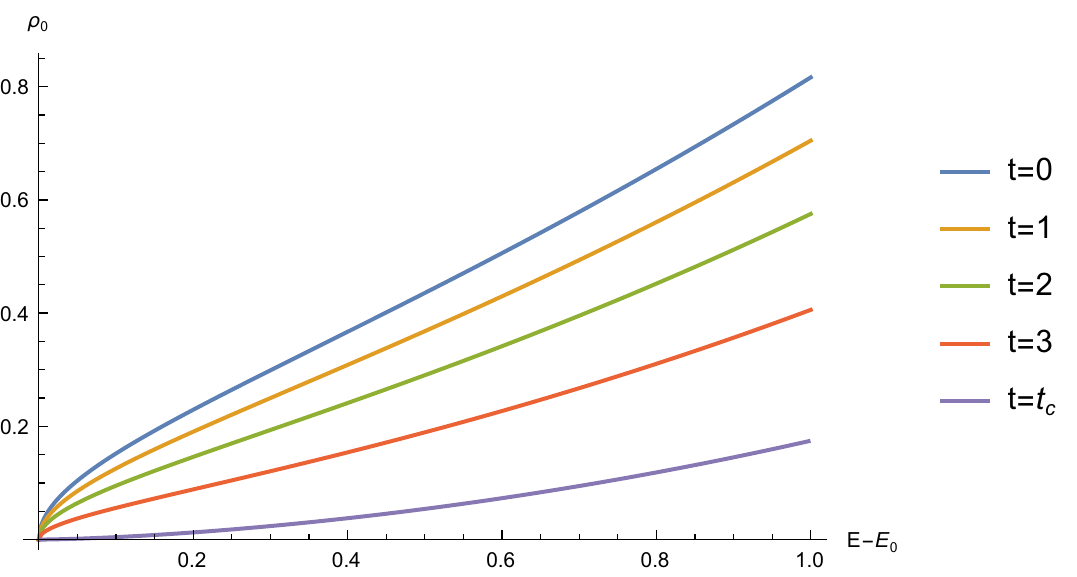}}
  \caption{
Plot 
of $\rho_0(E)$ 
against \subref{sfig:rho0} $E$ and \subref{sfig:rho0-shift} $E-E_0$
for anti-FZZT branes.
We set $\xi=18$ in this plot.
}
  \label{fig:rho0}
\end{figure}
Let us now focus on the case of JT gravity with anti-FZZT branes.
In Figure~\ref{fig:rho0} we plot $\rho_0(E)$ in \eqref{eq:rhoanti}
for several different values of $t$.
Due to property~1 of $S_{\text{therm}}$,
it is convenient to plot $\rho_0(E)$ against
$E-E_0$ (see Figure~\ref{sfig:rho0-shift})
to consider the behavior of the entropy.
Then we observe that the overall scale of $\rho_0$ clearly decreases
as $t$ grows.\footnote{As we see in \eqref{eq:D-rho},
$D(\cE)$ is proportional to $\rho_0(E)$ 
up to a prefactor. Since this prefactor is independent of
$t$, the graph of $D(\cE)$ against $\cE-\cE_0$ also decreases
as $t$ grows, in a similar way as $\rho_0(E)$ in
Figure~\ref{sfig:rho0-shift}.}
By crude approximation, the overall scale of $\rho_0$
gives that of $\til{Z}_n$ and thus
this implies the decreasing behavior of
$S_{\text{therm}}$, as explained in property~2.
More precisely, as described in property~3,
$S_{\text{therm}}$ is related to $\til{Z}_n$ by \eqref{eq:S-tZrel}.
We found numerically that each individual terms $\log\til{Z}_n$
and $n\bra \cE\ket$ are not necessarily monotonically decreasing
functions of $t$ for generic values of $\xi$,
but the sum of them is always monotonically decreasing.

To summarize, we have seen that the monotonically decreasing behavior
of $\tS_n$ is equivalent to that of the ``thermodynamic entropy''
$S_{\text{therm}}$
if we regard the replica index $n$ as the inverse temperature.
Its decreasing behavior is intuitively understood
from that of the overall scale of $\rho_0(E)$.

It is well known that the entropy of the Hawking radiation
is bounded from above by the thermodynamic entropy $S_{\text{BH}}$
of the black hole,
which is given by the area of horizon in the semi-classical
approximation. For an evaporating black hole, the area of horizon
decreases as time passes and this explains the decreasing behavior
of the Page curve.
As we saw in \eqref{eq:Sn-SBH},
$\tS_n$ (not $S_{\text{therm}}$) corresponds to 
$S_{\text{BH}}$ for $n>1$. 
In general, $\tS_n$ and $S_{\text{therm}}$
are different quantities.
However, one can easily see that $\tS_n$ becomes equal to $S_{\text{therm}}$
in the limit $n\to1$
\begin{equation}
\begin{aligned}
 \lim_{n\to1}\tS_n=\lim_{n\to1}S_{\text{therm}}.
\end{aligned} 
\end{equation}
Thus the thermodynamic entropy $S_{\text{BH}}$ of black hole literally
corresponds to the ``thermodynamic entropy'' $S_{\text{therm}}$
in the limit $n\to1$.

\section{Dilaton gravity}\label{sec:dilaton}

Recently, dilaton gravities with nontrivial dilaton potential
were studied as deformations of JT gravity
\cite{Maxfield:2020ale,Witten:2020wvy}.
Black hole solutions
in these gravities were also discussed in \cite{Witten:2020ert}.
JT gravity with (anti-)FZZT branes can be viewed as
this type of dilaton gravity \cite{Okuyama:2021eju}.
In this section we will study black hole solutions 
from the viewpoint of dilaton gravity.\footnote{See also
\cite{Gregori:2021tvs} for recent related studies.}

The action of dilaton gravity is written as 
\cite{Witten:2020ert}
\begin{equation}
\begin{aligned}
 I=-\hf\int d^2x\rt{g}(\phi R+W(\phi)).
\end{aligned} 
\end{equation}
We derived that in the case of JT gravity with $K$ (anti-)FZZT branes
the dilaton potential is given by \cite{Okuyama:2021eju}
\begin{equation}
W(\phi)=\left\{
\begin{aligned}
& 2\phi+\frac{t\sqrt{2\xi}}{\xi+\pi^2\phi^2}e^{-2\pi\phi}
&\quad&(\text{anti-FZZT}),\\
& 2\phi-\frac{t\sqrt{2\xi}}{\xi+\pi^2\phi^2}e^{-2\pi\phi}
&\quad&(\text{FZZT}).
\end{aligned} \right.
\label{eq:W-phi}
\end{equation}
The general Euclidean black hole solution is given by
\begin{equation}
\begin{aligned}
 ds^2=A(r)dt^2+\frac{dr^2}{A(r)},\quad \phi(r)=r,\quad
A(r)=\int_{r_h}^r d\phi W(\phi),
\end{aligned} 
\label{eq:BHsol}
\end{equation}
where $r=r_h$ is the horizon at which $A(r)$ vanishes.
This is a one-parameter family of solutions parametrized by $r_h=\phi_h$.
The value of $\phi_h$ is not fixed by the equation of motion.

The entropy of this solution is given by
\begin{equation}
\begin{aligned}
 S=2\pi \phi_h+S_0.
\end{aligned} 
\label{eq:S-phih}
\end{equation}
The physical condition is $A(r)>0$ for $r>r_h$. 

\begin{figure}[t]
\centering
\subcaptionbox{$W(\phi)$ for FZZT branes\label{sfig:W-FZZT}}
{\includegraphics
[width=0.45\linewidth]{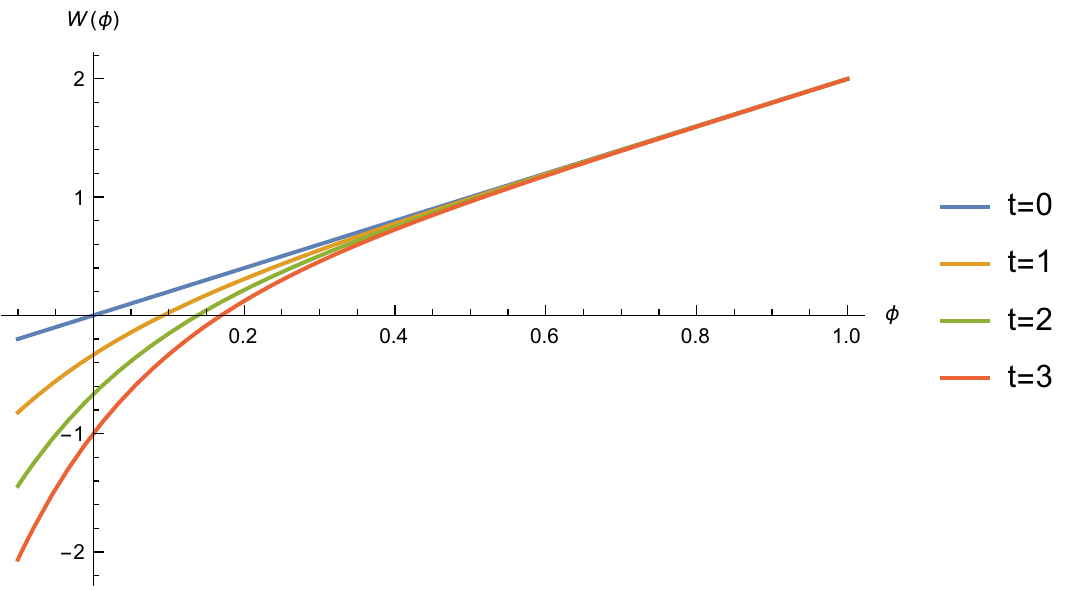}}
\hskip5mm
\subcaptionbox{$W(\phi)$ for anti-FZZT branes \label{sfig:W-antiFZZT}}
{\includegraphics
[width=0.45\linewidth]{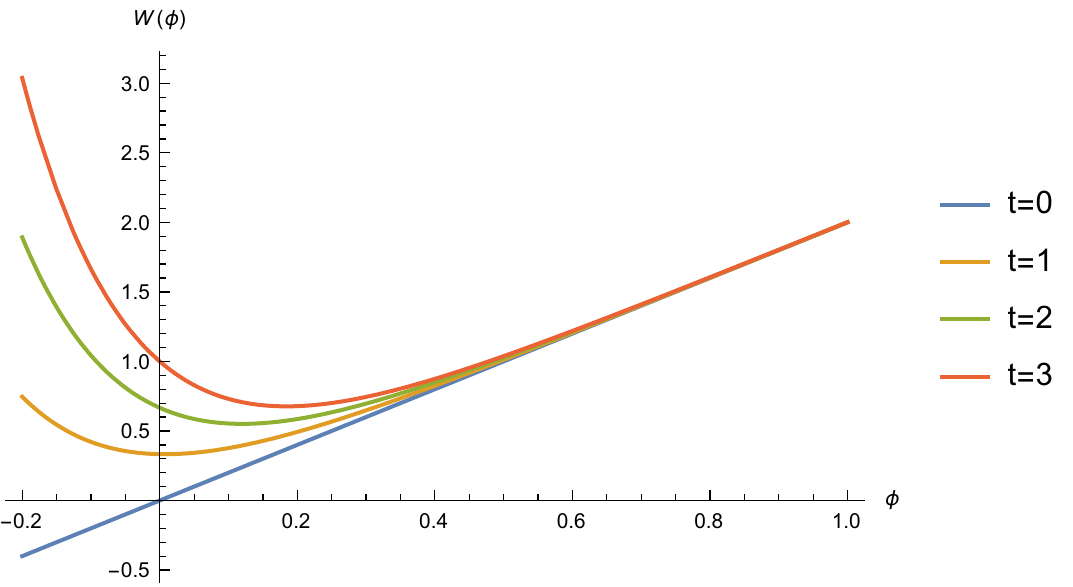}}
  \caption{
Plot 
of $W(\phi)$ in \eqref{eq:W-phi} 
for \subref{sfig:W-FZZT} FZZT branes and \subref{sfig:W-antiFZZT} 
anti-FZZT branes.
We set $\xi=18$ in this plot.
}
  \label{fig:W}
\end{figure}

For a fixed temperature $T$, $\phi_h$ is determined by the condition
\begin{equation}
\begin{aligned}
 \frac{W(\phi_h)}{4\pi}=T.
\end{aligned} 
\label{eq:W-T}
\end{equation}
In Figure \ref{fig:W} we show the plot of $W(\phi)$.
For the FZZT branes in Figure \ref{sfig:W-FZZT},
\eqref{eq:W-T} has a unique solution $\phi_h$ for a given value of $T$. 
As $t$ increases, $\phi_h$ also increases. 
Thus the entropy increases as a function of $t$.

\begin{figure}[t]
\centering
\[
\begin{tikzpicture}[scale=0.6]
\draw [thick] (0,1.36) -- (10,1.36); 
\draw [black,thick]  (0,1.8) .. controls (2,-1) and (4,1) .. (10,4);
\draw [black!75,thick]  (0,4) .. controls (2,0) and (4,1) .. (10,4);
\draw [black!65,thick]  (0,7) .. controls (2,1) and (4,1) .. (10,4);
\draw (10,1.36) node [right]{$4\pi T$};
\draw (10,4) node [right]{$W(\phi)$};
\draw (.2,1.8) node [text=black,above left]{$t<\tc$};
\draw (.2,4) node [text=black!75,above left]{$t=\tc$};
\draw (.2,7) node [text=black!65,above left]{$t>\tc$};
\coordinate (A) at (4.955,1.36);
\fill (A) circle (3pt);
\draw (5.2,1.36) node [below]{$\phi_h$};
\end{tikzpicture}
\]
\vspace{-8ex}
\caption{Stability of black hole solutions
in the anti-FZZT brane setup.
The largest root $\phi_h$ of \eqref{eq:W-T}
corresponds to the stable solution.
The solution no longer exists for $t>\tc$.}
\label{fig:BHcond}
\end{figure}
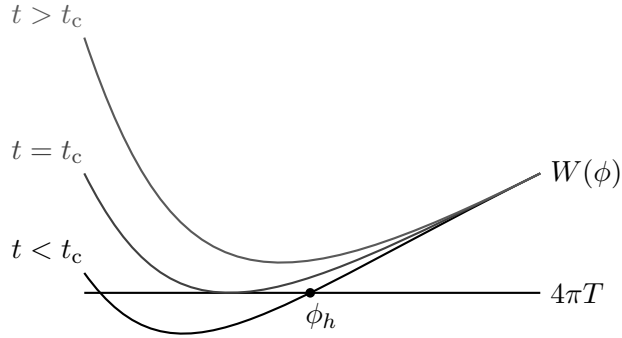

On the other hand, from Figure \ref{sfig:W-antiFZZT} 
for the anti-FZZT branes, 
one can see that there are two solutions of 
\eqref{eq:W-T} if $t$ is not too large.
As discussed in \cite{Witten:2020ert},
the stable solution with minimal free energy corresponds
to the largest root $\phi_h$ of \eqref{eq:W-T} 
(see Figure~\ref{fig:BHcond}).
The largest root $\phi_h$ of \eqref{eq:W-T} decreases as 
$t$ increases. 
This explains the decreasing behavior of
the entropy \eqref{eq:S-phih}.
Beyond some critical value $t=\tc$,
there is no solution of \eqref{eq:W-T} for a given temperature.
This might be interpreted that for $t>\tc$ 
there is no stable black hole solutions;
at $t=\tc$ the stable black hole disappears.
This suggests that,
to model the black hole evaporation process, the anti-FZZT brane
setup is more suitable than the FZZT brane setup.

\section{Conclusion and outlook}\label{sec:conclusion}

In this paper we studied the entanglement entropy
in the matrix model of JT gravity with anti-FZZT branes,
which serves as a toy model of an evaporating black hole.
The entanglement entropy is defined between
the color and flavor sectors,
which correspond respectively to bulk gravity and
to the interior partners of the early Hawking radiation.
We computed the entropy in the planar approximation
as well as in the 't Hooft limit.
The 't Hooft coupling $t$, which is proportional
to the number of branes, plays the role of time.
We computed numerically the von Neumann and R\'{e}nyi entropies
as functions of $t$.
In both cases, the entropy first increases and then decreases,
which is peculiar to the Page curve of an evaporating black hole.
We stress that
we treated the anti-FZZT branes as dynamical objects
and this was crucial to reproduce the late-time decreasing
behavior of the entropy, because otherwise the entropy
approaches a constant value at late time in the probe brane approximation
\cite{Penington:2019kki},
as we saw in Figure~\ref{fig:S-antiFZZT}.

We saw that the system exhibits a phase transition at $t=\tc$.
This may be viewed as the end of the evaporation of the black hole.
We studied the critical behavior of the entropy
and derived that it scales as in \eqref{eq:Seffbehavior}.
As $t$ grows toward $t=\tc$, the R\'{e}nyi entropy becomes dominated
by $\tS_n$ in \eqref{eq:tSn}.
We conjectured that $\tS_n$ monotonically decreases
and proved this conjecture in the large $\xi$ limit.
We also gave an intuitive explanation of this decreasing behavior.
We studied black hole solutions of dilaton gravity
that describes JT gravity with (anti-)FZZT branes
and saw the continuous growth of the entropy
in the FZZT setup as well as a phase transition in the anti-FZZT setup.
This suggests that the anti-FZZT brane
setup is more suitable to model an evaporating black hole.

There are many interesting open questions.
We have seen that our model of dynamical branes in JT gravity
serves as a good toy model for an evaporating black hole.
We hope that the behavior of our model beyond the phase transition $t>\tc$
would shed light on the deep question of 
the unitarity in black hole evaporation, e.g.~the
final state proposal in \cite{Horowitz:2003he}.
It would be interesting to study our matrix model 
beyond the phase transition $t>\tc$ along the lines of \cite{Gao:2021uro}.
Our analysis of the Page curve was limited to the planar approximation.
It would be interesting to compute the higher genus corrections to
the Page curve. More ambitiously, it would be very interesting if
we can compute the Page curve of our model non-perturbatively in $\gs$.
We leave this as an interesting future problem.
We can also repeat the analysis of the Petz map in \cite{Penington:2019kki}
using our setup of dynamical branes.
It would be interesting to study how the entanglement wedge reconstruction
is modified from the result of \cite{Penington:2019kki} if we
take account of the back-reaction of branes. 

In our calculation of the Page curve, the decreasing behavior of entropy 
comes from the
last term of \eqref{eq:Tr-rho^n}, which is interpreted as a 
contribution of replica wormholes \cite{Penington:2019kki,Almheiri:2019qdq}.
The appearance of the replica wormhole
is closely related to the ensemble average 
on the boundary side of the AdS/CFT correspondence.
The role of the ensemble average in the gravitational path integral
is still not well-understood and there are many conceptual issues
related to this problem, such as the factorization puzzle
(see e.g.~\cite{Marolf:2020xie,McNamara:2020uza,Saad:2021rcu,
Saad:2021uzi,Blommaert:2021fob,Heckman:2021vzx} and references therein).
It is believed that the R\'{e}nyi entropy is a self-averaging quantity 
\cite{Penington:2019kki}.
Nonetheless, it would be interesting to see how the Page curve of our
model would look like if we pick a certain member of the ensemble and 
do not take an average over the random matrix
(see e.g.~\cite{Blommaert:2021etf} for a study in this direction).

\acknowledgments
This work was supported in part by JSPS KAKENHI Grant
Nos.~19K03845, 19K03856, 21H05187
and JSPS Japan-Russia Research Cooperative Program.
A preliminary result of this work was presented by one of the authors (KO)
in the KMI colloquium at Nagoya University on October 13, 2021.

\appendix

\section{Schwinger-Dyson equation from saddle point approximation}
\label{app:SDeq}

In this appendix we will derive the Schwinger-Dyson equation
\eqref{eq:SD-R} based on the saddle point method.

Let us consider the integral
\begin{equation}
\begin{aligned}
 \int \prod_{i=1}^Kd\phi^\dag_id\phi_i
 e^{-\sum_{ij}\phi_i^\dag(\la\cob_{ij}-\varrho_{ij})\phi_j}.
\end{aligned} 
\end{equation}
Then the two point function $\b{\phi_i\phi_j^\dag}$ is equal to the resolvent
\begin{equation}
\begin{aligned}
 \b{\phi_i\phi_j^\dag}=(\la-\varrho)^{-1}_{ij}.
\end{aligned} 
\end{equation}
We can rewrite the integral as
\begin{equation}
\begin{aligned}
 \int d\phi^\dag d\phi e^{-\phi^\dag(\la-\varrho)\phi}&=
\int d\phi^\dag d\phi dG_{ij}\cob(G_{ij}-\phi_i\phi_j^\dag)
 e^{-\Tr (\la-\varrho)G}\\
&=\int d\phi^\dag d\phi dGd\Si e^{\Si_{ji}(G_{ij}-\phi_i\phi_j^\dag)
 -\Tr (\la-\varrho)G}\\
&=\int dGd\Si e^{\Tr \Si G-\Tr\log \Si-\Tr (\la-\varrho)G}.
\end{aligned} 
\label{eq:GSi-int}
\end{equation}
The density matrix $\varrho_{ij}$ is given by
\begin{equation}
\begin{aligned}
 \varrho_{ij}=\frac{C_i^\dag A C_j}{KZ_1}=C_i^\dag \h{A} C_j,\qquad
\h{A}=\frac{A}{KZ_1}.
\end{aligned} 
\end{equation}
After integrating out $C^\dag,C$ we have
\begin{equation}
\begin{aligned}
 \int dC^\dag dC e^{-\Tr C^\dag C+\Tr \varrho G}=e^{-\Tr\log (1-\h{A}G)},
\end{aligned} 
\end{equation}
where $\Tr\log (1-\h{A}G)$ should be understood as
the trace of both color and flavor indices.
Then \eqref{eq:GSi-int} becomes $\int dGd\Si e^{-I}$
where the action $I$ is given by
\begin{equation}
\begin{aligned}
 I=-\Tr \Si G+\Tr\log\Si+\la\Tr G+\Tr\log(1-\h{A}G).
\end{aligned} 
\end{equation}
In the planar approximation, the $G$- and $\Si$-integrals
can be evaluated by
the saddle point approximation. The saddle point equations read
\begin{equation}
\begin{aligned}
 \frac{\del I}{\del\Si_{ij}}&=-G_{ij}+(\Si^{-1})_{ij}=0,\\
 \frac{\del I}{\del G_{ij}}
 &=-\Si_{ij}+\la\cob_{ij}-\cob_{ij}\Tr\frac{\h{A}}{1-\h{A}G}=0.
\end{aligned} 
\label{eq:del-I}
\end{equation}
Multiplying the second equation of \eqref{eq:del-I} by $G_{ij}$
and summing over $i,j$, we find
\begin{equation}
\begin{aligned}
 -K+\la\Tr G-\Tr G\Tr\frac{\h{A}}{1-\h{A}G}=0.
\end{aligned} 
\end{equation}
This is rewritten as
\begin{equation}
\begin{aligned}
 \la\Tr G&=K+\Tr G\Tr\frac{\h{A}}{1-\h{A}G}\\
&=K+\Tr G\sum_{n=1}^\infty \Tr\h{A}^n\Tr G^{n-1}.
\end{aligned} 
\label{eq:saddle}
\end{equation}
In the planar approximation we have
\begin{equation}
\begin{aligned}
\Tr G&\approx\sum_i\b{\phi_i\phi_i^{\dag}}
 =\sum_i(\la-\varrho)^{-1}_{ii}=R(\la).
\end{aligned} 
\end{equation}
We also find
\begin{equation}
\begin{aligned}
 \Tr G^n&=\b{ \phi_{i_1} \phi^\dag_{i_2}\phi_{i_2} \phi^\dag_{i_3}\cdots  
\phi_{i_n}\phi^\dag_{i_1}}\\
&\approx \b{\phi^\dag_{i_1}\phi_{i_1}}
 \cdot\b{\phi^\dag_{i_2}\phi_{i_2}}\cdots
 \b{\phi_{i_n}^\dag \phi_{i_n}}\\
&=(\Tr G)^n=R(\la)^n.
\end{aligned} 
\end{equation}
Finally, \eqref{eq:saddle} becomes
\begin{equation}
\begin{aligned}
 \la R(\la)=K+\sum_{n=1}^\infty\Tr \h{A}^n R(\la)^n.
\end{aligned} 
\end{equation}
Thus we have re-derived the Schwinger-Dyson equation \eqref{eq:SD-R},
which was originally derived
by means of diagrams in \cite{Penington:2019kki}.
The above saddle point method can be generalized
to the Grassmann-odd integral (i.e.~to the case of FZZT branes).

\bibliography{paper}
\bibliographystyle{utphys}

\end{document}